\documentclass[twocolumn]{aastex62}

\newcommand{\cnjwl}{$cn_{\rm JWL}$}
\newcommand{\cnjwlcor}{$cn_{\rm JWL,cor}$}

\newcommand{\cacn}{Ca--CN}
\newcommand{\cnw}{CN-w}
\newcommand{\cns}{CN-s}
\newcommand{\vvhb}{$V - V_{\rm HB}$}

\newcommand{\vhb}{$V_{\rm HB}$}
\newcommand{\vbump}{$V_{\rm bump}$}

\newcommand{\kms}{km s$^{-1}$}

\newcommand{\nrgb}{$n$(\cnw):$n$(\cns)}
\newcommand{\nfergb}{$n$(N-normal):$n$(N-enhanced)}

\newcommand{\agbrgb}{$n$(AGB)/$n$(RGB)}

\newcommand{\vby}{$(b-y)$ versus $V$}

\newcommand{\vcn}{\cnjwlcor\ versus $V$}

\newcommand{\hst}{{\it HST}}

\newcommand{\sdss}{SDSS}

\newcommand{\cnwave}{$\lambda$3883}
\newcommand{\scn}{$S$(3839)}
\newcommand{\ds}{$\delta S$(3839)}
\newcommand{\dy}{$\Delta Y$}
\newcommand{\dcy}{$\delta cy$}
\newcommand{\dmo}{$\delta m1$}

\newcommand{\dc}{$\Delta_{\rm F275W,F814W}$}
\newcommand{\dtrio}{$\Delta_{\rm C~F275W,F336W,F438W}$}

\newcommand{\agbm}{AGB-manqu\'e}

\newcommand{\vrot}{$v_{\rm rot}$}
\newcommand{\arot}{$A_{\rm rot}$}

\newcommand{\dlam}{$\Delta\lambda$}
\newcommand{\mavg}{$\langle$\cnjwlcor$\rangle_{\rm 25\ pts}$}

\shorttitle{NGC 6752}
\shortauthors{Lee}

\begin{document}

\title{MULTIPLE STELLAR POPULATIONS OF GLOBULAR CLUSTERS FROM HOMOGENEOUS \cacn\ PHOTOMETRY. III. NGC 6752.
\footnote{Based on observations made with the Cerro Tololo Inter-American Observatory (CTIO) 1 m telescope, which is operated by the SMARTS consortium.}}

\author[0000-0002-2122-3030]{Jae-Woo Lee}
\affiliation{Department of Physics and Astronomy, Sejong University\\
209 Neungdong-ro, Gwangjin-Gu, Seoul, 05006, Korea\\
jaewoolee@sejong.ac.kr, jaewoolee@sejong.edu}

\begin{abstract}
We present a multiple stellar population (MSP) study of the globular cluster (GC) NGC~6752. We show that our new photometric CN index accurately traces the CN and the nitrogen abundances in cool giants, finding the discrete double red giant branch (RGB) and asymptotic giant branch (AGB) sequences with number ratios between the CN-weak and the CN-strong populations of \nrgb\ = 25:75 ($\pm$3; RGB) and 79:21 ($\pm$13; AGB). The discrepancy in these number ratios suggests that a significant fraction of the low-mass \cns\ stars failed to evolve into the AGB phase. However, unlike previous studies, our results indicate the presence of an extreme \cns\ AGB population in NGC~6752, which may require follow-up spectroscopic study. Similar to what is seen for M5, the evolution of the nitrogen abundance is discrete and discontinuous, while the evolutions of oxygen and sodium are continuous between the two populations in NGC~6752, implying that different astrophysical sources are responsible for the evolutions of these elements. In addition, the helium abundance inferred from the RGB bump magnitude shows hat the \cns\ population is slightly more helium-enhanced. Despite the identical cumulative radial distributions between the two populations, the structure-kinematics coupling can be observed in individual populations: the \cnw\ population has a spatially elongated shape with a faster rotation, while the \cns\ population shows weak or no net rotation, with spatially symmetric shape, raising important question about the long-term dynamical evolution of the GCs.
\end{abstract}


\section{INTRODUCTION}
Over the past decade, it becomes very clear that the classical paradigm in which GCs are composed of a simple stellar population is no longer valid.
The longstanding problems of the ubiquitous nature of the bimodal CN abundance distributions and the Na--O anticorrelations are good examples of the MSPs in GCs \citep[e.g.,][]{smith87,carretta09}.
How GC systems form and evolve are currently unknown and understanding the true nature of GCs is an ongoing goal for near-field cosmology \citep[e.g.,][]{jwlnat,lee15,lee17,piotto15,renzini15}.

It is believed that the significant spreads or the bimodal distributions of the lighter elemental abundances in GC stars result from the chemical pollution by the first generation (FG) of stars which had experienced proton-capture processes at high temperatures\footnote{For example, see \citet{lee10} for the potential effects of the internal mixing in the GC RGB stars.} \citep[e.g.,][]{dercole08,carretta09}.
However, how exactly normal GCs formed in the early phase of the Milky Way is still under debate and the astrophysical sources of such chemical pollution remain to be unknown.

It is well known that the variations in the lighter elemental abundances, such as C, N, and O, can greatly affect the ultraviolet (UV) and the visual passbands through  OH, NH, CN, CH, and C$_2$ in the atmospheres of cool giant stars in GCs.
These molecular absorption bands are very strong, so that even broadband photometry in the UV can detect variations of the lighter elemental abundances.
In this regard, photometry is still very important for the sake of ease and completeness, particularly in the very crowded central part of GCs where traditional spectroscopy cannot be applied.\footnote{See \citet{lee16} for the non-trivial aspects of the LTE analysis of high-resolution spectroscopy of GC RGB stars.}
When studying the MSPs in GCs, therefore, spectroscopy and photometry are complementary \citep[see, for example,][]{lee17}.

In our previous study, we pointed out the potential problems of a MSP study of GCs based on the broadband UV photometry (\dlam\ $>$ 50 nm), such as the \hst\ magic-trio, Johnson $U$, \sdss\ $u$ and Washington $C$.
The photometric indices from these broadband systems cover numerous very strong absorption features such as, OH, NH, CN, CH, C$_2$, Mg II h and k, and Ca II H and K, in their passbands.
Furthermore, each element listed above has a different degree of the luminosity effects and elemental enhancement and depletion among different populations.
As a consequence, the interpretation of the photometric products from these systems can be somewhat complicated and ambiguous.
In sharp contrast, our \cnjwl\ index (= $JWL39 - Ca_{\rm new}$) is specifically designed to measure the CN \cnwave\ molecular band absorption strength only.
Therefore, our \cnjwl\ index is as good as traditional spectroscopic indices such as \scn\ and \ds, and is capable of distinguishing MSPs in normal GCs with great satisfaction.
On top of that, since it is a photometric index, our new approach easily be applied to the dense central part of GCs.

The spatial distributions and the kinematic properties of individual populations in GCs from a large field of view (FOV) are also important ingredients for understanding the formation and the evolution of GCs.
As some recent numerical simulations have suggested \citep[see, for example,][]{vesperini13}, statistics from sufficiently large FOVs is essential to grasp an accurate view of the MSPs in GCs.
In our previous study of M22 and M5 \citep{lee15,lee17}, we showed that the MSPs in these GCs show a structure-kinematics coupling, i.e., the structural parameters of individual populations are closely linked with their kinematic properties,  such as rotation.
Therefore, the spatial distributions and the kinematic properties of individual populations can provide crucial information on the formation of MSPs in GCs, and, at the same time, address some important questions regarding the long-term dynamical evolution of the Galactic GC systems.

Finally, our new system is perfectly suited to study the AGB stars. 
Since the effective temperatures of AGB stars are cool enough to maintain the CN molecular band formation in the atmospheres, we can detect AGB stars in a large FOV, which is very critical for performing a complete census of AGB populations.

This is a part of a series of papers addressing the MSPs of Galactic GCs based on our homogeneous \cacn\ photometry.
Over the last decade, we have devoted tremendous effort to developing a new photometric system, which has allowed us to measure the heavy and CN abundances simultaneously.
In this study, we investigate the MSPs of the RGB and the AGB stars in NGC~6752 using our new filter system.

\begin{deluxetable*}{cccccccccc}
\tablenum{1}
\tablecaption{Integration times (s) \label{tab:obs}}
\tablewidth{0pc}
\tablehead{
\multicolumn{5}{c}{CTIO Filters} &
\multicolumn{1}{c}{} & \multicolumn{4}{c}{New Filters} \\
\cline{1-5}\cline{7-10}
\colhead{$y$} & \colhead{$b$} & \colhead{$v$} & \colhead{$u$} & 
\colhead{$Ca$} & \colhead{} & 
\colhead{$y$} &
\colhead{$b$} & 
\colhead{$Ca_{\rm new}$} &
\colhead{$JWL39$} 
}
\startdata
2990 & 6300 & 6000 & 7300 & 16300 & & 3875 & 9240 & 31900 & 11000 \\ 
\enddata 
\end{deluxetable*}

\section{OBSERVATIONS AND DATA REDUCTION}\label{s:reduction}
We observed NGC~6752 for 33 nights, 15 of which were photometric, in 10 runs from 2008 August to 2014 May using the CTIO 1.0 m telescope. 
The CTIO 1.0 m telescope was equipped with a STA 4k $\times$ 4k CCD camera, providing a plate scale of 0\farcs289 pixel$^{-1}$ and a FOV of about 20\arcmin\ $\times$ 20\arcmin.
A detailed discussion of our new filter system can be found in  \citet{lee15,lee17}. 
The average seeing for the full observation period was 1\farcs37 $\pm$ 0\farcs20.
The total integration times for NGC~6752 are shown in Table~\ref{tab:obs}.

The raw data handling is described in detail in our previous works \citep{n6723,lee15,lp16,lee17}.
The photometry of NGC~6752 and standard stars was analyzed using the DAOPHOTII, DAOGROW, ALLSTAR and ALLFRAME, and  COLLECT-CCDAVE-NEWTRIAL packages \citep{pbs87,pbs94,pbs95,lc99}.

\begin{figure}
\epsscale{1.0}
\figurenum{1}
\plotone{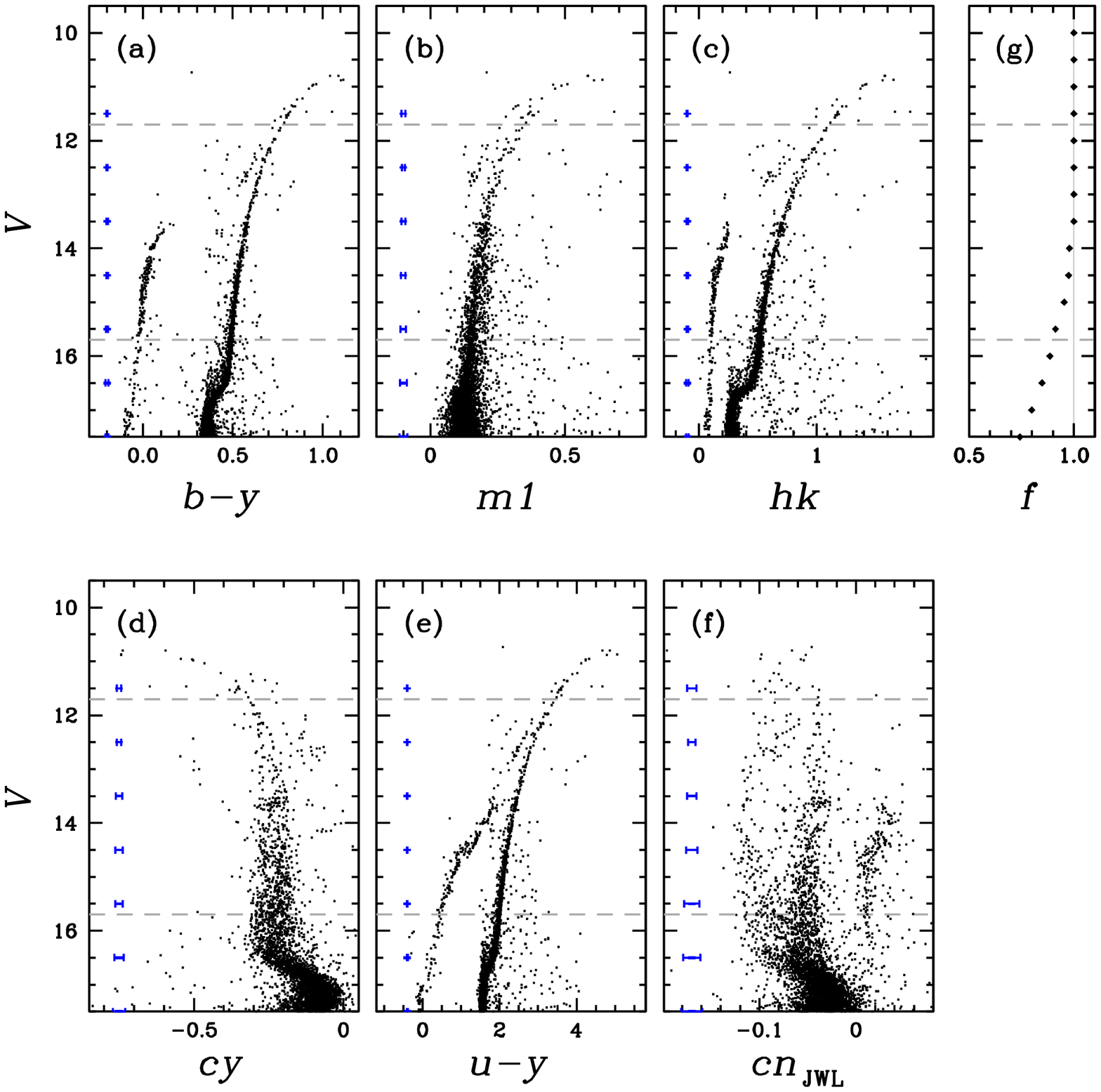}
\caption{(a)--(f) CMDs of NGC~6752. We show only the good quality stars using the separation index \citep{pbs03}.
The error bars denote the mean measurement errors at given magnitude bins.
The dashed horizontal lines are for \vvhb\ = $\pm$2.0 mag, with \vhb(NGC~6752) = 13.70 mag \citep{harris96}.
The discrete double-RGB sequences are evident in the \cnjwl\ CMD.
(g) Completeness fractions against $V$ magnitude.
}\label{fig:cmd}
\end{figure}

As discussed in \citet{lee17}, we employed a new strategy for the ground-based observations to make use of the positional information from the \hst\ photometry by \citet{anderson08}.
Through this process the detection rate in the central part of the cluster can be greatly improved. Our ground-based photometry is complete at $V \approx$ 13.75 mag and the detection fraction at $V$ = 15.5 mag becomes 95.4\% as shown in Figure~\ref{fig:cmd}(g).
Due to the unavoidable seeing effects in ground-based observations, our photometry is slightly incomplete at the magnitude range of our interest, $-2 \leq$ \vvhb\ $\leq$ 2 mag,  where \vhb\ is the $V$ magnitude of the horizontal branch (HB), in the central part of the cluster.
The undetected stars from our ground-based observations are located within $r \lesssim 1.5 r_c$\footnote{The core and the half-light radii of NGC~6752 are $r_c$ $\approx$ 10\arcsec\ and $r_h$ $\approx$ 115\arcsec\ \citep{harris96}.} from the center.
However, we will show later that the incomplete detection in the central part of the cluster does not affect our results presented here.
The effective FOV of our NGC~6752 science field was about 35\arcmin\ $\times$ 25\arcmin\ and the total number of stars measured from our ALLFRAME run was about 50,000.

Finally, the astrometric solutions for individual stars were calculated using the data extracted from the Naval Observatory Merged Astrometric Dataset \citep[NOMAD,][]{nomad} and the IRAF IMCOORS package.

\begin{figure}
\epsscale{1.0}
\figurenum{2}
\plotone{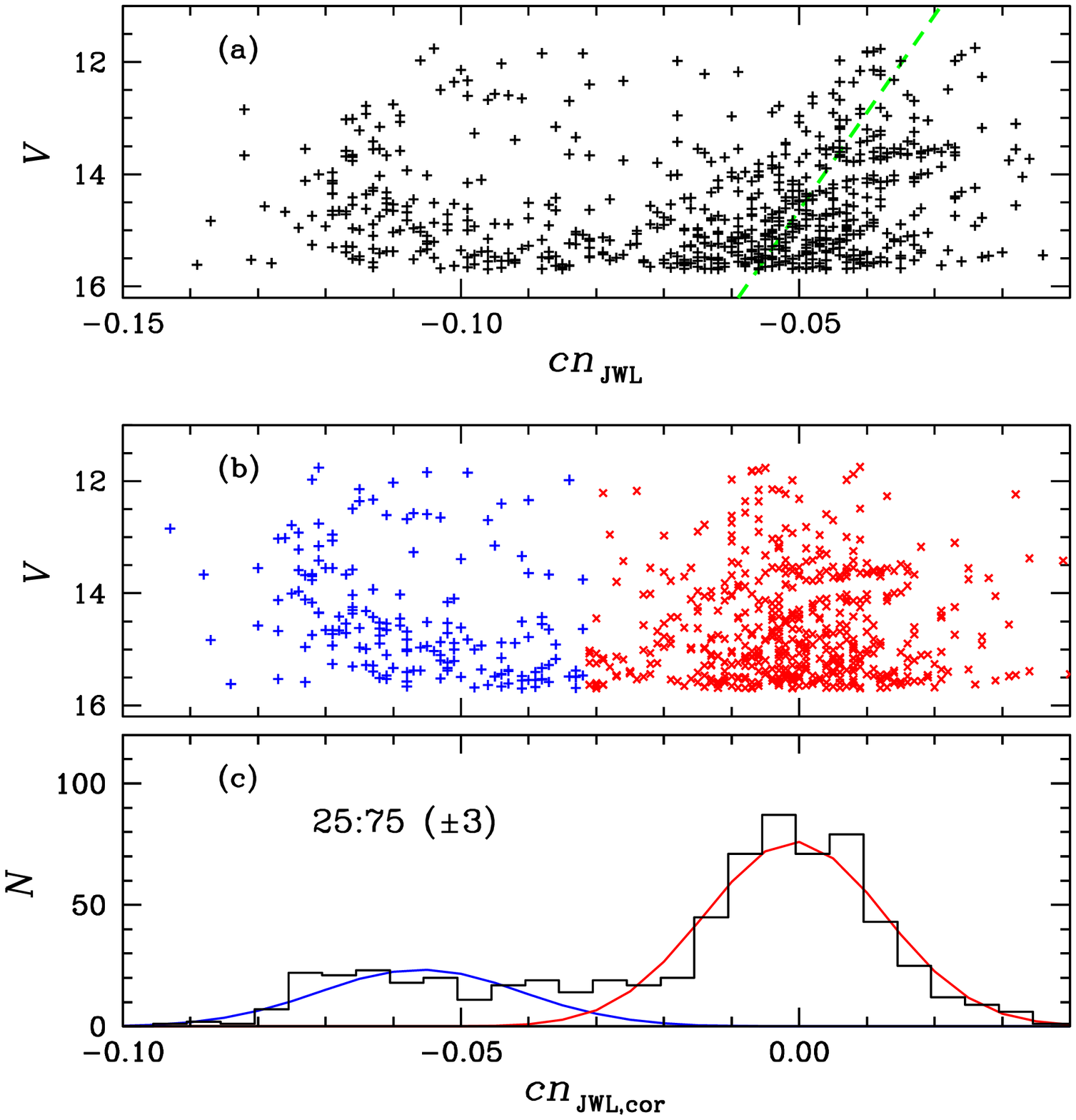}
\caption{(a) A \cnjwl\ versus $V$ CMD for all RGB stars with $-2 \leq$ \vvhb\ $\leq$ 2.0 mag, showing a weak luminosity effect.
The green dotted line denotes the linear fit to the \cns\ RGB sequence.
(b) A \vcn\ CMD. The blue plus signs denote the \cnw\ and the red crosses denotes the \cns\ RGB stars.
(c) Histogram of the RGB stars and the two Gaussian fitted lines.
}\label{fig:rgbcmd}
\end{figure}

\begin{deluxetable*}{crrrrrr}
\tablecaption{Coefficients and the goodness of the fit between $S(3839)$ [and \ds] by \citet{norris81} and \cnjwlcor, $cy$ and $m1$
\label{tab:deltas}}
\tablenum{2}
\tablewidth{0pc}
\tablehead{
\multicolumn{2}{c}{Index} &
\multicolumn{1}{c}{Slope\tablenotemark{a}} &
\multicolumn{1}{c}{Intercept\tablenotemark{a}}&
\multicolumn{1}{c}{$p$-value}&
\multicolumn{1}{c}{$\rho$\tablenotemark{b}}
}
\startdata
\cnjwlcor & \scn  & 2.394 $\pm$ 0.534 & 0.365 $\pm$ 0.024 & 0.000 & 0.556 \\
       & \ds   & 4.080 $\pm$ 0.383 & 0.368 $\pm$ 0.017 & 0.000 & 0.846 \\
&&&&& \\
$m1$   & \scn  & 0.358 $\pm$ 0.241 & 0.200 $\pm$ 0.071 & 0.144 & 0.216 \\
       & \ds   & $-$0.688 $\pm$ 0.257 & 0.450 $\pm$ 0.076 & 0.010 & $-$0.371 \\
&&&&& \\
\dmo\tablenotemark{1}
        & \scn  & 0.758 $\pm$ 0.498 & 0.292 $\pm$ 0.024 & 0.136 & 0.221 \\
        & \ds   & $-$0.476 $\pm$ 0.568 & 0.263 $\pm$ 0.027 & 0.406 & $-$0.124 \\
&&&&& \\
$cy$   & \scn  & $-$0.021 $\pm$ 0.213 & 0.295 $\pm$ 0.066 & 0.923 & $-$0.015 \\
       & \ds   & 0.738 $\pm$ 0.212 & 0.471 $\pm$ 0.066 & 0.001 & 0.461 \\
&&&&& \\
\dcy\tablenotemark{2}
       & \scn  & 0.086 $\pm$ 0.275 & 0.302 $\pm$ 0.024 & 0.757 & 0.046 \\
       & \ds   & 0.832 $\pm$ 0.282 & 0.272 $\pm$ 0.025 & 0.005 & 0.402 \\
\enddata
\tablenotetext{a}{$S(3839)$ or \ds\ = Intercept + Slope $\times$ Index.}
\tablenotetext{b}{Pearson's correlation coefficient.}
\tablenotetext{1}{\dmo\ = $m1 - (3.034 - 0.368V + 0.012V^2$)}
\tablenotetext{2}{\dcy\ = $cy - (-2.663 + 0.333V - 0.011V^2$)}
\end{deluxetable*}

\section{RESULTS}

\subsection{Color--magnitude diagrams}\label{s:cmd}
We show the color-magnitude diagrams (CMDs) of NGC~6752 in Figure~\ref{fig:cmd}.
Our $(b-y)$, $hk$ [= (Ca$_{\rm new} - b$) $-$ ($b-y$)] CMDs show very narrow RGB sequences for NGC~6752, reflecting the monometallic nature of the cluster, consistent with the results from previous high-resolution spectroscopy by others \citep[see, for example][]{yong13}.
On the other hand, very broad or even discrete double RGB sequences can be seen in $m1$, $cy$, and \cnjwl\ CMDs, which are strong evidence of the variation of the lighter elemental abundances.
It will be discussed later that the nitrogen abundance measurements from the NH molecular band at $\lambda$ 3360\AA\ by \citet{yong08} exhibit a bimodal distribution, consistent with the discrete double RGB sequence in our \cnjwl\ index.

Based on our multi-color photometry, we chose the membership RGB stars in NGC~6752.
In our previous work \citep{lee15,lee17}, we showed that our approach is very effective for selecting off-cluster populations in GCs.
Since NGC~6752 is located at a rather higher Galactic latitude ($b \approx -26^{\circ}$), the field star contamination in NGC~6752 is expected to not be severe in our results.
Also, the magnitude range of interest to us is 1 -- 2 magnitude brighter than the ranges of M22 and M5, therefore the number of off-cluster field stars should be much smaller in NGC~6752.

In Figure~\ref{fig:rgbcmd}(a), our \cnjwl\ index for RGB stars shows a weak gradient against the $V$ magnitude, indicative of the existence of differential effective temperature effect and differential surface gravity effect on the CN molecular band formation.
Therefore, we derived a linear fit along the \cns\ RGB stars (see below for the definition) and we obtained the following relation to correct the luminosity effect in our \cnjwl\ index:
\begin{equation}
 cn_{\rm JWL, cor} = cn_{\rm JWL} - (3.420\times10^{-2} - 5.705\times10^{-3}V.)
\end{equation}
We show the corrected CMD of the bright RGB stars in panel (b), where the gradient against the $V$ magnitude is effectively removed.

The number ratio between the two groups of RGB stars in NGC~6752 was derived using the expectation maximization (EM) algorithm for the two-component Gaussian mixture model.
Note that the \cnw\ population is defined to be stars with smaller \cnjwlcor\ values at given $V$ magnitudes, while the \cns\ population is those with larger \cnjwlcor\ values.
Stars with $P$(\cnw$|x_i) \geq$ 0.5 from the EM estimator calculations represent the \cnw\ population, where $x_i$ denotes the individual RGB stars, while those with $P$(\cns$|x_i)$ $>$ 0.5  correspond to the \cns\ population. 
Our calculations showed that the number ratio between the two populations is \nrgb\ = 25:75 ($\pm$3), consistent with that of \citet{milone17}, who obtained \nrgb\ = 29:71 ($\pm$ 2) in the central part of the cluster ($\lesssim$ 0.7$r_h$), within the statistical errors.

\begin{figure}
\epsscale{1.0}
\figurenum{3}
\plotone{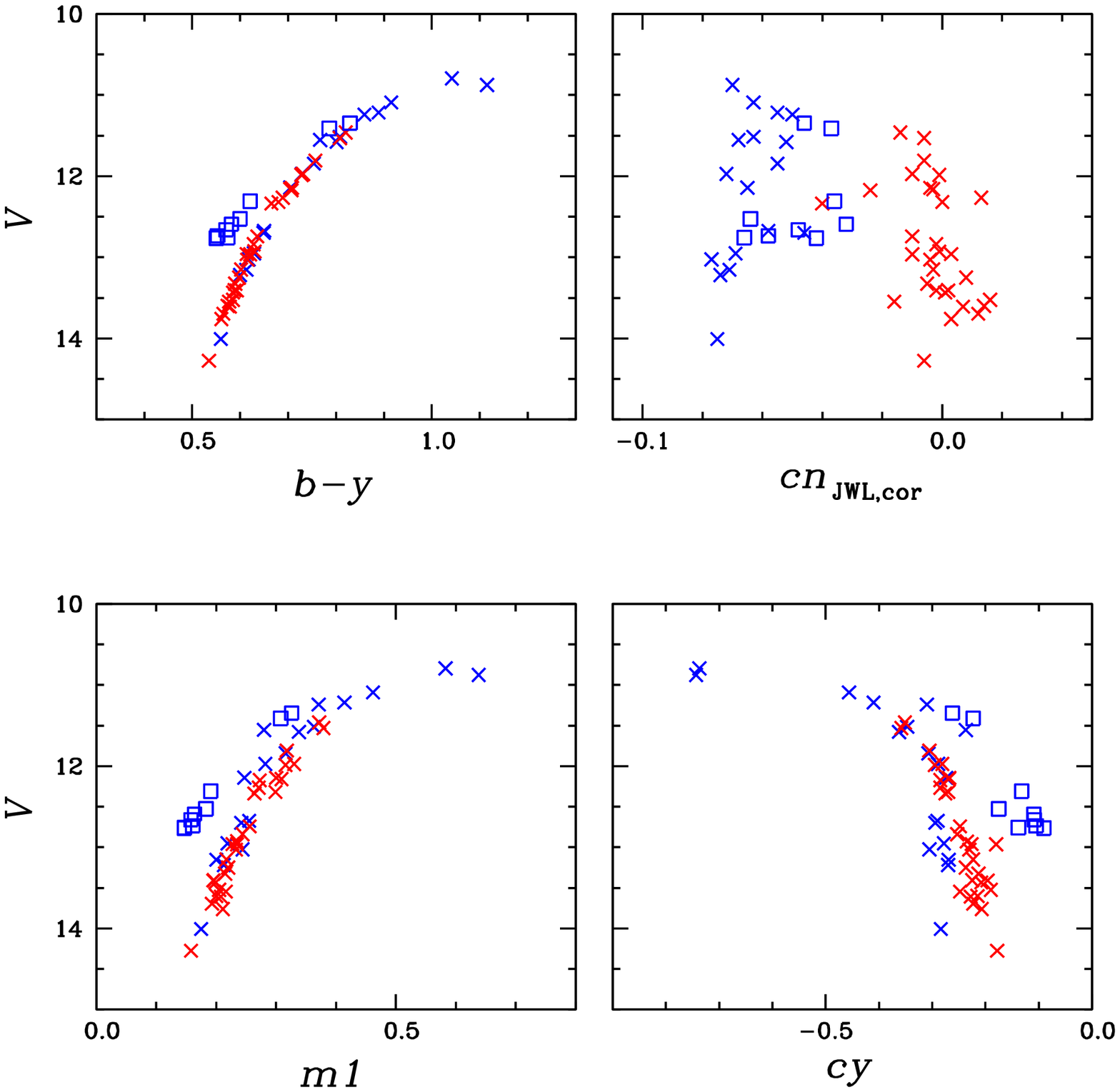}
\caption{CMDs for the stars studied by \citet{norris81}.
The crosses are for the RGB stars and the open squares are for the AGB stars.
The blue color denotes the \cnw\ population, while the red color denotes the \cns\ population classified by \citet{norris81} based on their \scn\ measurements.
Note that no \cns\ AGB stars was reported by \citet{norris81}.
The discrete double RGB sequences are evident in the \vcn\ CMD.
}\label{fig:cmdnorris}
\end{figure}

\begin{figure}
\epsscale{1.0}
\figurenum{4}
\plotone{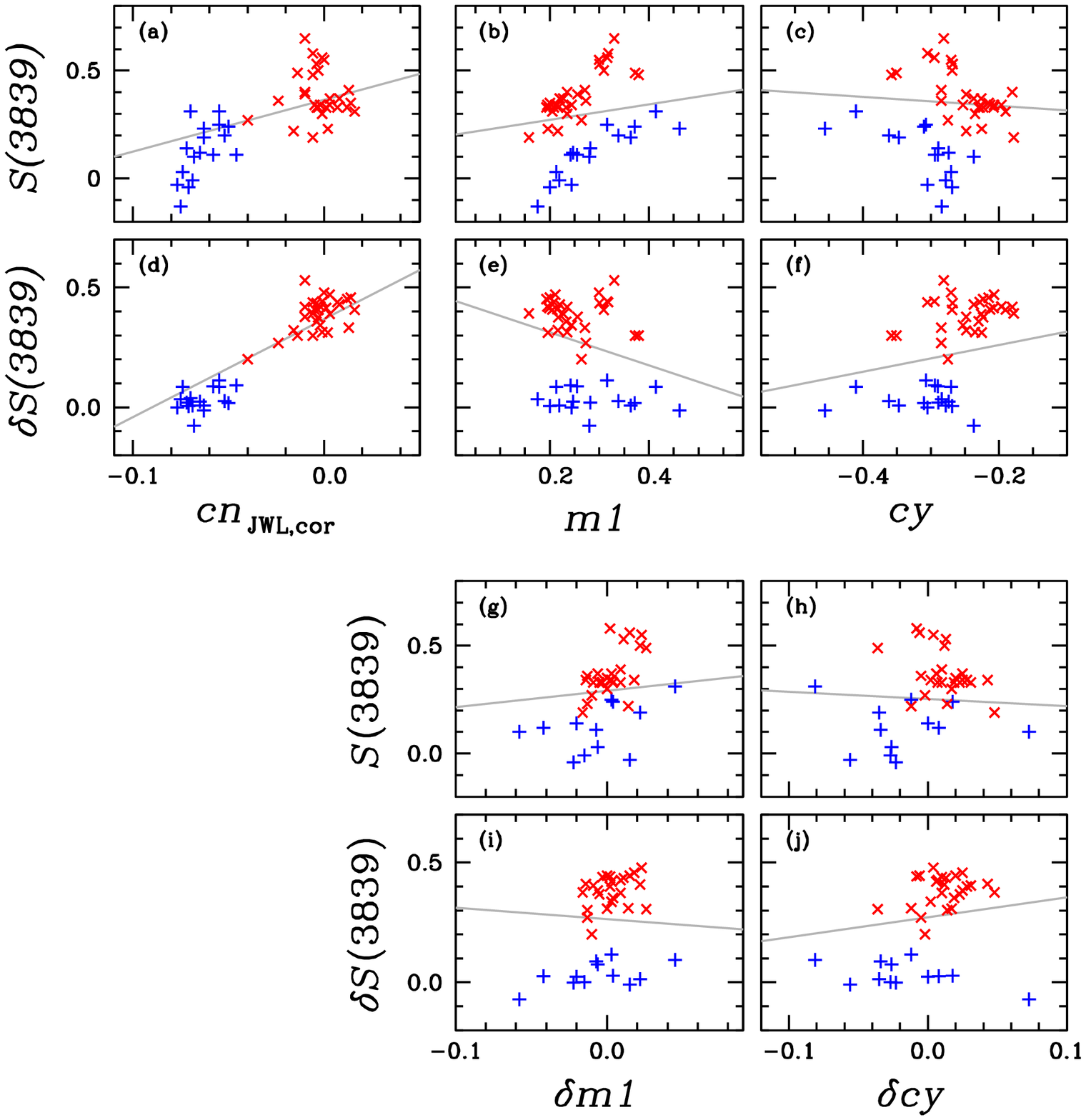}
\caption{(a)--(c) Correlations between selected color indices and the CN \cnwave\ strength, \scn.
The blue color denotes the \cnw\ and the red color denotes the \cns\ RGB stars classified by \citet{norris81}. 
The gray solid lines are linear fits to the data.
(d)--(f) Correlations between selected color indices and the CN excess,\ds. 
Note that the linear fit between the \cnjwlcor\ and \ds\ is greatly improved.
(g)--(j) Correlations between $\delta m1$ and $\delta cy$ and the \scn\ and \ds.
Note that correcting luminosity effects on the $m1$ and the $cy$ indices does not improve the fits.
}\label{fig:deltas}
\end{figure}

\subsection{Red giant stars}
\subsubsection{A comparison with \citet{norris81}}
In their pioneering study, \citet{norris81} performed a low-resolution spectroscopic study of NGC~6752 RGB and AGB stars and they measured some spectral indices including \scn, the CN \cnwave\ molecular band absorption strengths.
Figure~\ref{fig:cmdnorris} shows our CMDs for the giant stars that \citet{norris81} studied.
As shown in the figure, the separation between the \cnw\ and the \cns\ by \citet{norris81} can be clearly seen in the \vcn\ CMD, while severe confusion makes it impossible to distinguish the two stellar populations in other color indices, including the $m1$ and the $cy$.
To quantitatively examine the utility of individual color indices as CN tracers, we investigated the correlations between  the \scn by \citet{norris81} and the color indices presented in this study.
We show plots of \scn\ versus individual color indices in Figure~\ref{fig:deltas}(a)--(c), and the goodness of the fits in Table~\ref{tab:deltas}.
It is generally believed that $m1$ and $cy$ indices can measure CN and NH abundances;  however, the \cnw\ and the \cns\ stars classified by \citet{norris81} are superposed onto each other on the $m1$ and the $cy$ CMDs, suggesting that both the $m1$ and the $cy$ are less efficient than \cnjwl\ to trace the CN variations.
This was also noted in our previous study of M5 \citep{lee17}.

In Figure~\ref{fig:deltas}(a), our \cnjwlcor\ index can nicely separate the two populations.
However, the scatter around the fitted line is rather large, due to the luminosity effect on \scn.
We derived the CN excess, \ds, which is defined to be the distance in \scn\ from the lower envelope of RGB stars in the $V$ magnitude versus \scn, to correct the luminosity effect, 
\begin{equation}
\delta S = S(3839) - (1.682 - 0.131V).
\end{equation}
Note that the relation given by \citet{norris81} is different from ours,\footnote{The relation given by \citet{norris81} was
\begin{equation}
\delta S = S(3839) - ( 1.832 - 0.146V).
\end{equation}
}
due to slightly different $V$ magnitudes between the two studies.
Using our CN excess, we show our results in Figure~\ref{fig:deltas}(d)--(f).
The linear fit to our \cnjwlcor\ index can be greatly improved when we adopt \ds, with a correlation coefficient of 0.846 and $p$-value of 0.000 (see Table~\ref{tab:deltas}).
On the other hand, both the $cy$ and the $m1$ indices fail to correctly trace the CN excess.

As Figures~\ref{fig:cmd} and \ref{fig:cmdnorris} showed, the significant curvatures of the RGB sequences in the $m1$ and the $cy$ CMDs suggest the presence of the luminosity effect.
Therefore, in order to correct the luminosity effect on these two indices, we derived \dmo\ and \dcy, the excess in the $m1$ and $cy$ indices, finding
\begin{equation}
\delta cy = cy - (-2.030 + 0.244V - 0.008V^2)
\end{equation}
and
\begin{equation}
\delta m1 = m1 - (2.821 - 0.340V + 0.011V^2).
\end{equation}
We show our results in Figure~\ref{fig:deltas}(g--j).
As shown, both \dcy\ and \dmo\ do not mitigate the discrepancy, strongly suggesting that the poor fits in these two indices are not due to the luminosity effect but due to their ill nature as CN tracers.

\begin{figure}
\epsscale{1.0}
\figurenum{5}
\plotone{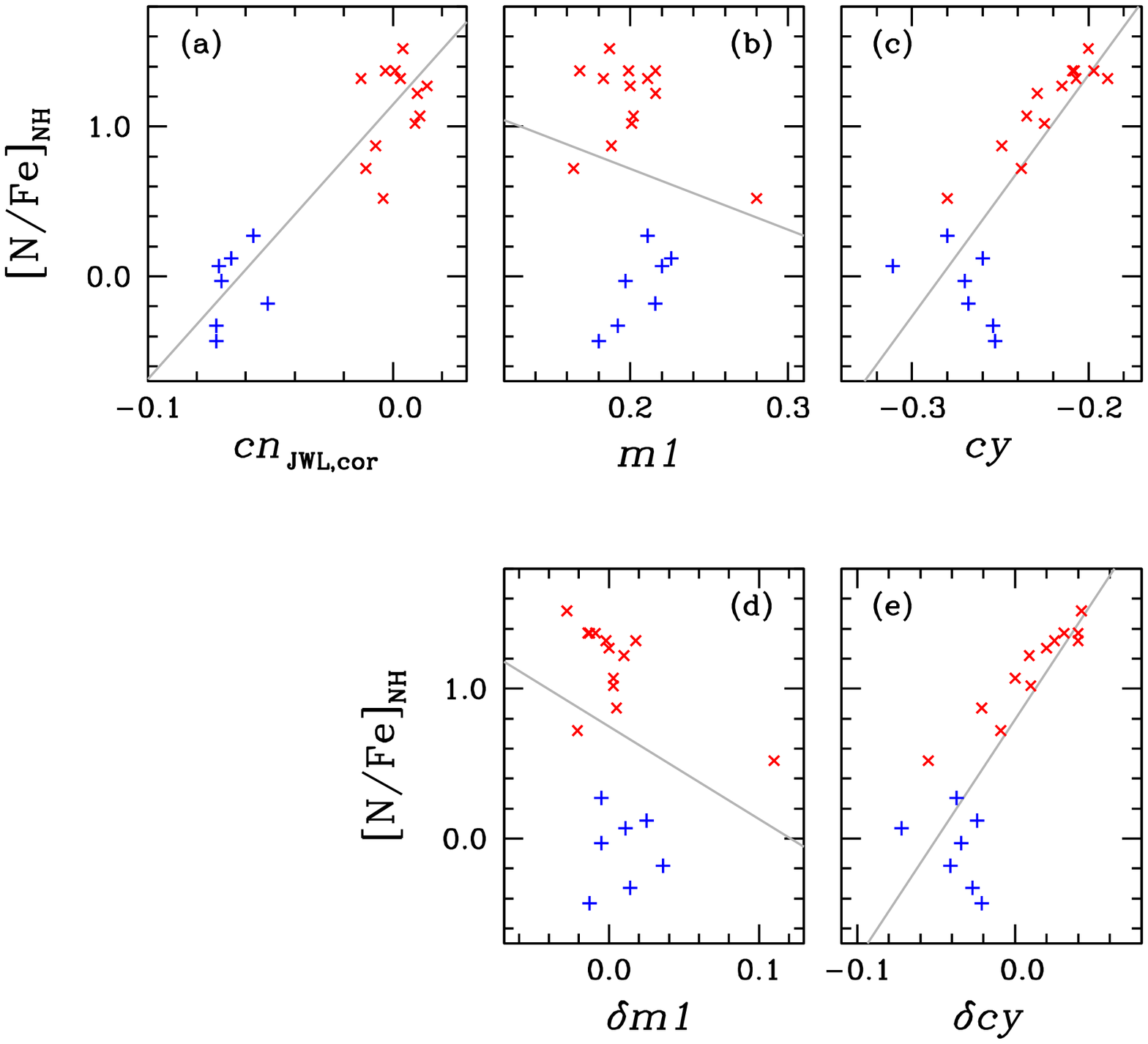}
\caption{Correlations between individual color indices and nitrogen abundances of NGC~6752 RGB stars \citep{yong08}.
The symbols are the same as those of the previous figures.
}\label{fig:cnvsphot}
\end{figure}

\begin{figure}
\epsscale{1.0}
\figurenum{6}
\plotone{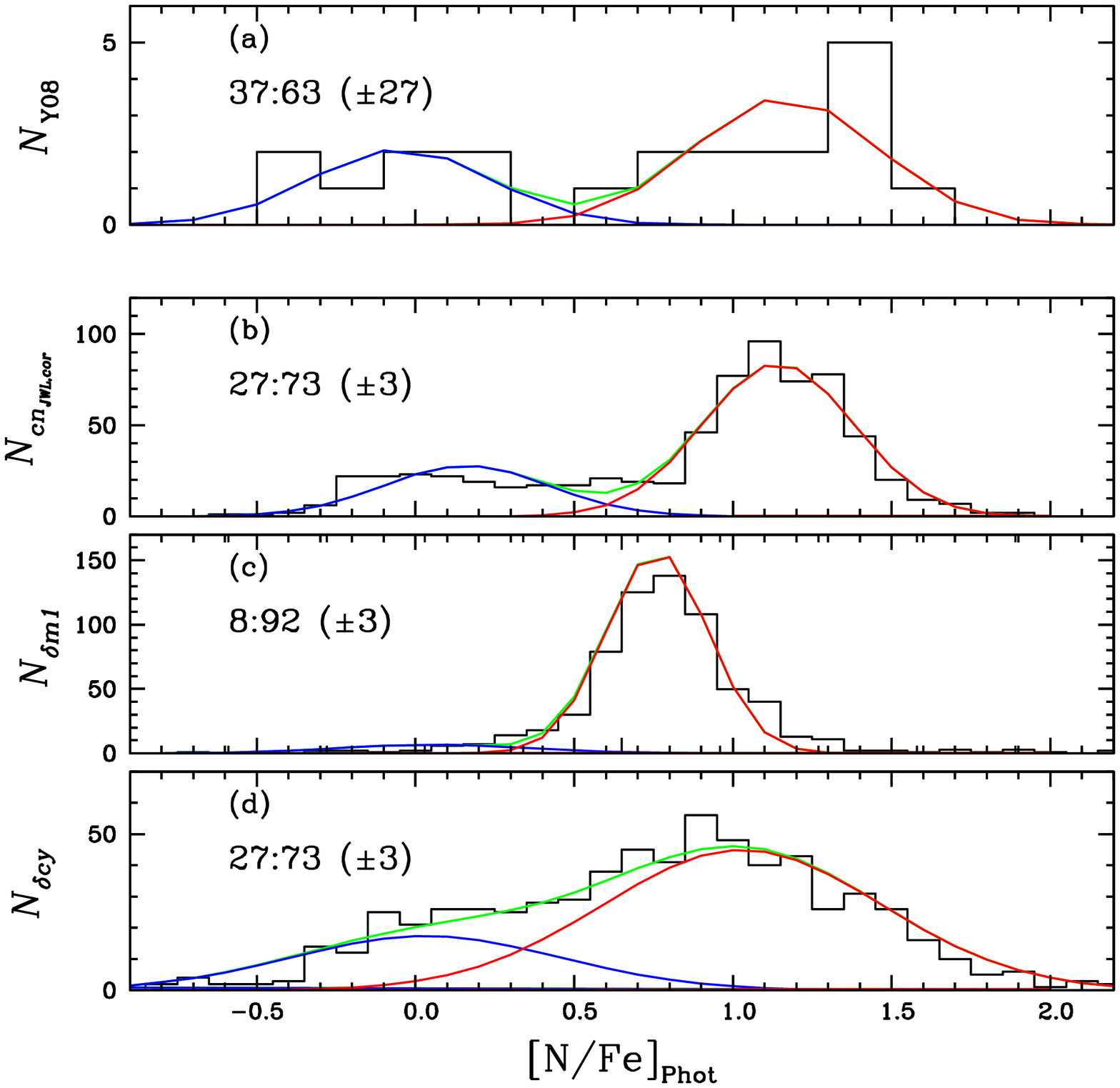}
\caption{(a) Histogram of the spectroscopic nitrogen abundance by \citet{yong08}.
(b) Photometric nitrogen abundance from our \cnjwlcor\ index using the relation given in Table~\ref{tab:fit}, which is in excellent agreement with that of \citet{yong08}.
(c) Photometric nitrogen abundance from \dmo\ index, which fails to reproduce that of \citet{yong08}.
(d) Photometric nitrogen abundance from \dcy\ index, which provides a comparable number density for the two RGB populations.
However, the dispersions of individual populations are very large and the separation between the two populations is not clearly defined.
}\label{fig:photN}
\end{figure}

\subsubsection{A comparison with \citet{yong08}}
Next, we explore the suitability of each color index for being a nitrogen abundance indicator.
It is well known that the GC giants show a positive CN--NH correlation and, hence, the CN molecular band strength can trace the nitrogen abundance.
In Figure~\ref{fig:cnvsphot} and Table~\ref{tab:fit}, we show least square fits for each color index and the nitrogen abundance from the NH band at $\lambda$ 3360 \AA\ \citep{yong08}.
The correlation between the nitrogen abundances and our \cnjwlcor\ index is excellent, with a $p$-value of 0.000 and correlation coefficient of 0.915.
The goodness of the fit for the nitrogen abundances and $cy$ and \dcy\ appears to be good numerically.
However, panels (c) and (e) reveal severe superposition of the two populations.

We derived the photometric nitrogen abundance for the individual RGB stars.
First, we performed Hartigan's dip-test, finding that the distribution of the nitrogen abundance of NGC~6752 RGB stars by \citet{yong08} is not unimodal, with $D$ = 0.058 and $p$-value = 0.899.
Therefore, assuming a bimodal nitrogen abundance distribution for NGC~6752 RGB stars, we applied the EM estimation for the two-component Gaussian mixture distribution model to derive the contributions from two RGB populations.
For each RGB star, we calculated the probability of it being a member of an N-normal or N-enhanced population.
Our result is shown in Figure~\ref{fig:photN}(a), with a number ratio between the N-normal and the N-enhanced populations of \nfergb\ = 37:63 ($\pm$27).
This population ratio is in agreement with those from photometric indices (see Figures~\ref{fig:rgbcmd} and \ref{fig:hst}) within the statistical error.

Using the least-square fits given in Table~\ref{tab:fit}, we calculated the photometric nitrogen abundances of the RGB stars with $-$2 $\leq$ \vvhb\ $\leq$ 2 mag.
The photometric nitrogen abundance from our \cnjwlcor\ index shows a well-defined bimodal distribution in Figure~\ref{fig:photN}(b).
The number ratio of the two populations is \nfergb\ = 27:73 ($\pm$3), consistent with that from spectroscopic measurements by \citet{yong08}.
Our result strongly supports the idea that our \cnjwlcor\ index is an excellent measure of the nitrogen abundance.
In sharp contrast, the photometric nitrogen abundance from the \dmo\ index shown in Figure~\ref{fig:photN}(c) does not agree with the nitrogen abundance measurements by \citet{yong08}, showing a conspicuous single peak with \nfergb\ = 8:92 ($\pm$3).
The population ratio from \dcy\ is \nfergb\ = 27:73 ($\pm$3) and is in agreement with that of \citet{yong08}.
However, the dispersion in the nitrogen abundance from \dcy\ in each population is twice as large as that from \citet{yong08} and no clear separation between the two populations can be found.

\begin{deluxetable*}{crrrrr}[t]
\tablecaption{Coefficients and the goodness of the fit between 
selected color indices and the nitrogen abundance by \citet{yong08}
\label{tab:fit}}
\tablenum{3}
\tablewidth{0pc}
\tablehead{
\multicolumn{1}{c}{Index} &
\multicolumn{4}{c}{[N/Fe]} \\
\cline{2-5}
\multicolumn{1}{c}{} &
\multicolumn{1}{c}{Slope\tablenotemark{a}} &
\multicolumn{1}{c}{Intercept\tablenotemark{a}}&
\multicolumn{1}{c}{$p$-value}&
\multicolumn{1}{c}{$\rho$\tablenotemark{b}}
}
\startdata
\cnjwlcor    &   18.312 $\pm$ 1.957 & 1.147 $\pm$ 0.079 & 0.000 &    0.915 \\
$m1$         & $-$4.071 $\pm$ 6.317 & 1.533 $\pm$ 1.291 & 0.528 & $-$0.154 \\ 
$\delta m1$  & $-$6.167 $\pm$ 5.280 & 0.746 $\pm$ 0.157 & 0.259 & $-$0.273 \\ 
$cy$         &   16.091 $\pm$ 2.820 & 4.562 $\pm$ 0.682 & 0.000 &    0.811 \\ 
$\delta cy$  &   15.990 $\pm$ 2.573 & 0.794 $\pm$ 0.089 & 0.000 &    0.833 \\ 
\enddata
\tablenotetext{a}{[N/Fe] = Intercept + Slope $\times$ color index.}
\tablenotetext{b}{Pearson's correlation coefficient.}
\end{deluxetable*}

\begin{figure}
\epsscale{1.0}
\figurenum{7}
\plotone{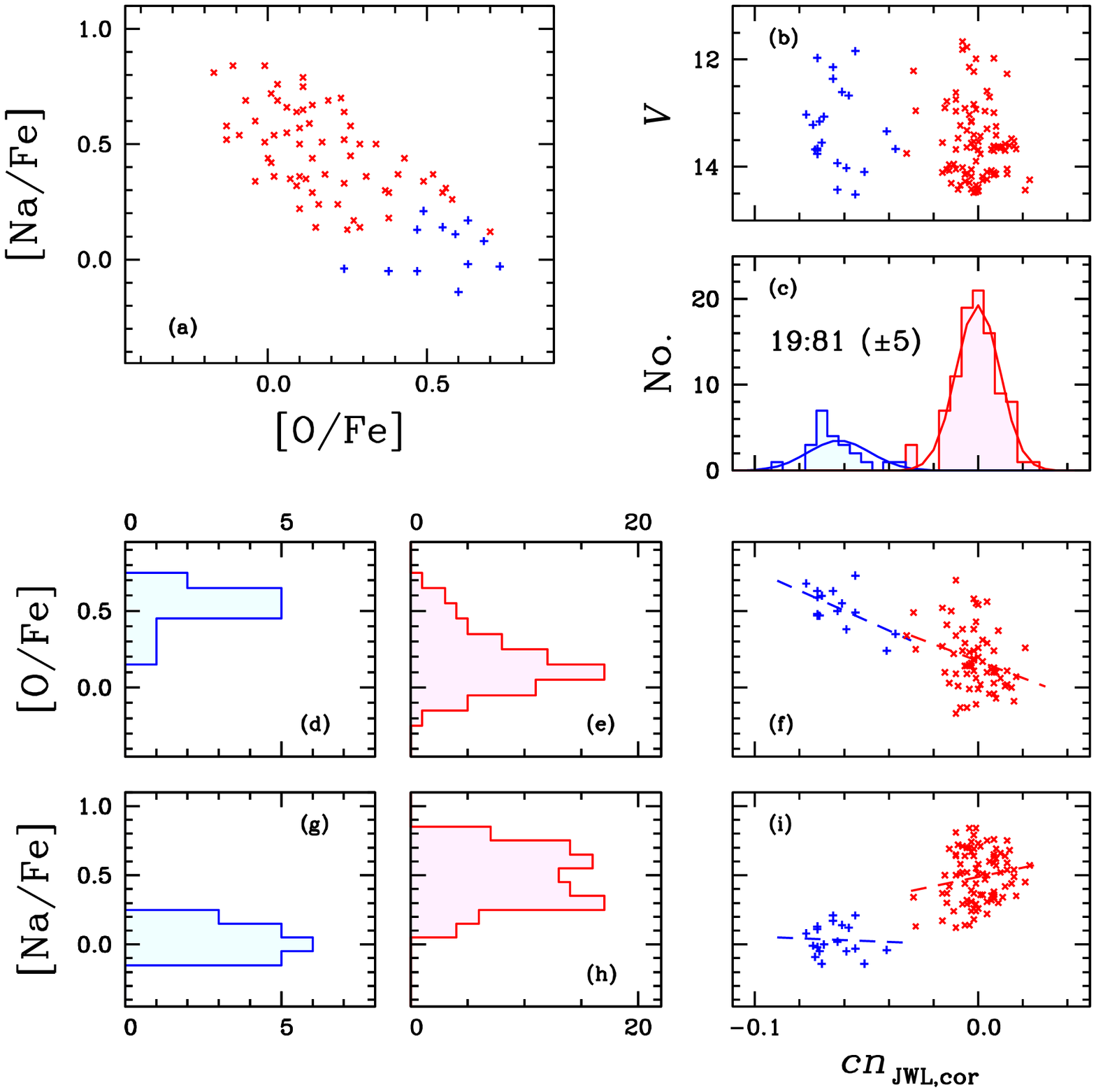}
\caption{(a) Na--O anticorrelation of NGC~6752 RGB stars in \citet{carretta07}.
The \cnw\ and \cns\ stars returned from the EM estimator are shown with blue plus signs and red crosses, respectively.
(b) \cnjwlcor\ versus $V$ CMD of RGB stars studied by \citet{carretta07}.
(c) \cnjwlcor\ distribution.
(d)--(e) [O/Fe] distributions of the \cnw\ and the \cns\ RGB stars.
(f) Plot of \cnjwlcor\ versus [O/Fe]. 
Note the clearly defined separate relations between \cnjwlcor\ and [O/Fe].
(g)--(h) [Na/Fe] distributions of the \cnw\ and the \cns\ RGB stars.
(i) Plot of \cnjwlcor\ versus [Na/Fe].
Similar to (f), there appear to exist two separate relations between \cnjwlcor\ and [Na/Fe], indicating that the transition between the two populations may not be continuous.
}\label{fig:nao}
\end{figure}
\begin{figure}
\epsscale{1.0}
\figurenum{8}
\plotone{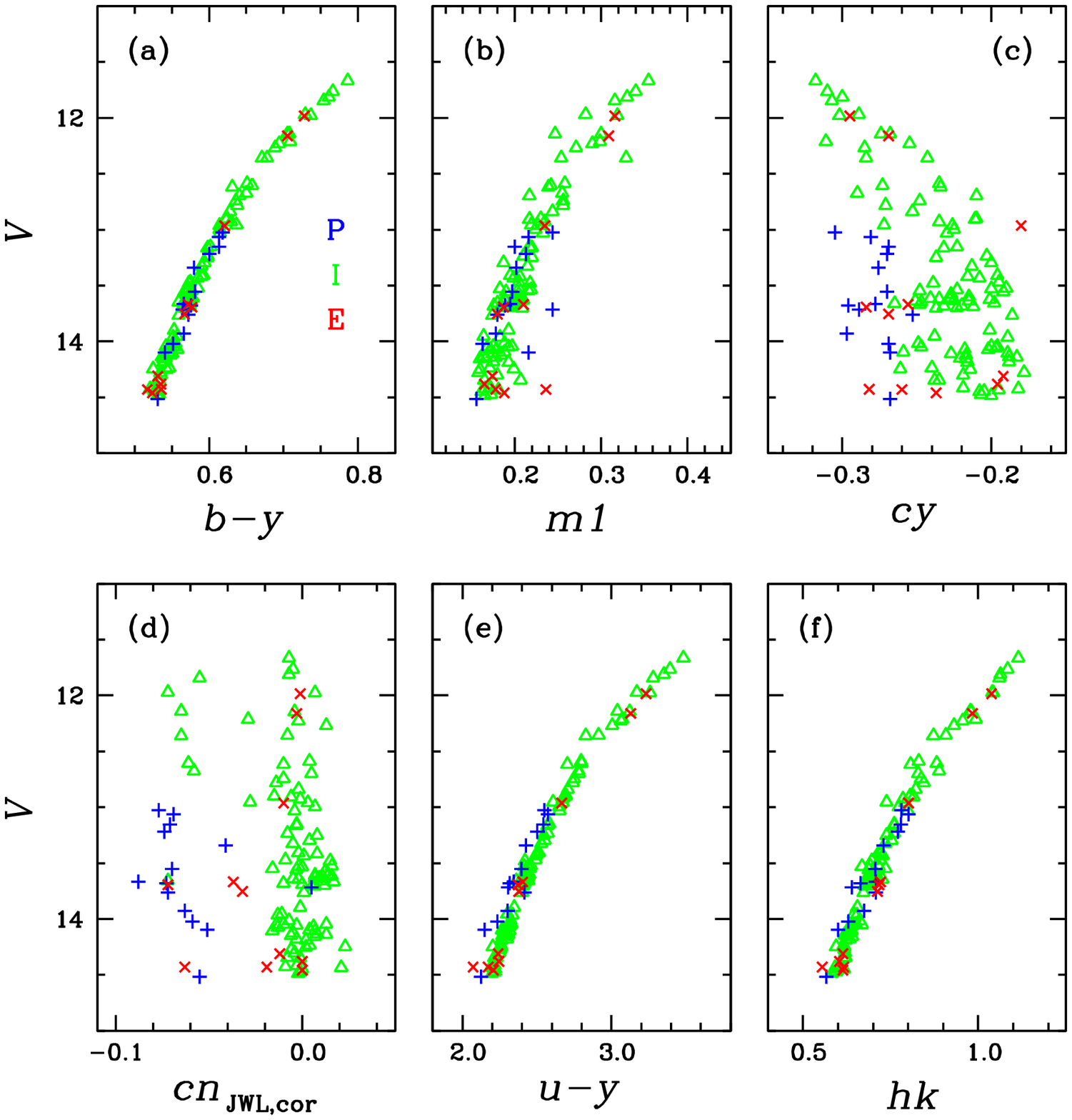}
\caption{CMDs of the RGB stars of \citet{carretta07}.
The blue plus sings denote the primordial population, the green open triangles denote the intermediate population, and the red crosses denote the extreme population defined by \citet{carretta07}.
It is clear that there is no difference between the intermediate and the extreme populations defined by \citet{carretta07} from the photometric point of view.
Also note that the primordial component defined by \citet{carretta07} does not occupy the region brighter than $V \approx$ 13 mag.
}\label{fig:pie}
\end{figure}

\subsubsection{A comparison with \citet{carretta09}}
In Figure~\ref{fig:nao}(a), we show the Na--O anticorrelation for the NGC~6752 RGB stars using the elemental abundance measurements by \citet{carretta07}.
As shown in Figure~\ref{fig:nao}(b), our \vcn\ CMD shows the double RGB sequences, with a number ratio of \nrgb\ = 19:81 ($\pm$5), marginally in agreement with our previous result, 25:75 ($\pm$3).
Figure~\ref{fig:nao}(d)-(i) show the [O/Fe] and the [Na/Fe] distributions for each population.
It is clear that our \cnjwlcor\ index distinguishes the primordial and other (i.e.\ the intermediate and the extreme) populations with great accuracy.
As shown, the \cnw\ population from our \cnjwlcor\ index has lower sodium and higher oxygen abundances, while the \cns\ population has higher sodium and lower oxygen abundances, consistent with the general trend in elemental abundances of the normal GCs.
It is imperative to note that there exist two separate \cnjwlcor--[O/Fe] and \cnjwlcor--[Na/Fe] relations, i.e., the two separate [N/Fe]--[O/Fe] and [N/Fe]--[Na/Fe] relations for both populations, which was already found in M5 \citep{lee17,smith13}.
It can be posited that the discontinuous evolution of the nitrogen abundance and continuous evolution of the oxygen and the sodium abundances suggest different astrophysical sources for these elements.

Finally, we show CMDs of RGB stars of \citet{carretta07} in Figure~\ref{fig:pie}.
The figure confirms our previous conclusion that both $m1$ and $cy$ are less efficient than our \cnjwl\ index for identifying the distinct populations of NGC~6752.
From our photometric point of view, no clear separation between the intermediate and the extreme populations can be seen. 
As we already pointed out in our previous work, it is likely due to the arbitrary definition of the boundary between the extreme and the intermediate populations in the continuous [O/Fe]--[Na/Fe] anticorrelation.
We also note that the lack of a primordial component in the brighter RGB stars with $V \lesssim$ 13 mag is evident.
The origin of the non-uniform distribution of the primordial component of NGC~6752 against the $V$ magnitude is not clear and it is beyond the scope of this study \citep[see also,][]{lee10}.

\begin{figure}
\epsscale{1.0}
\figurenum{9}
\plotone{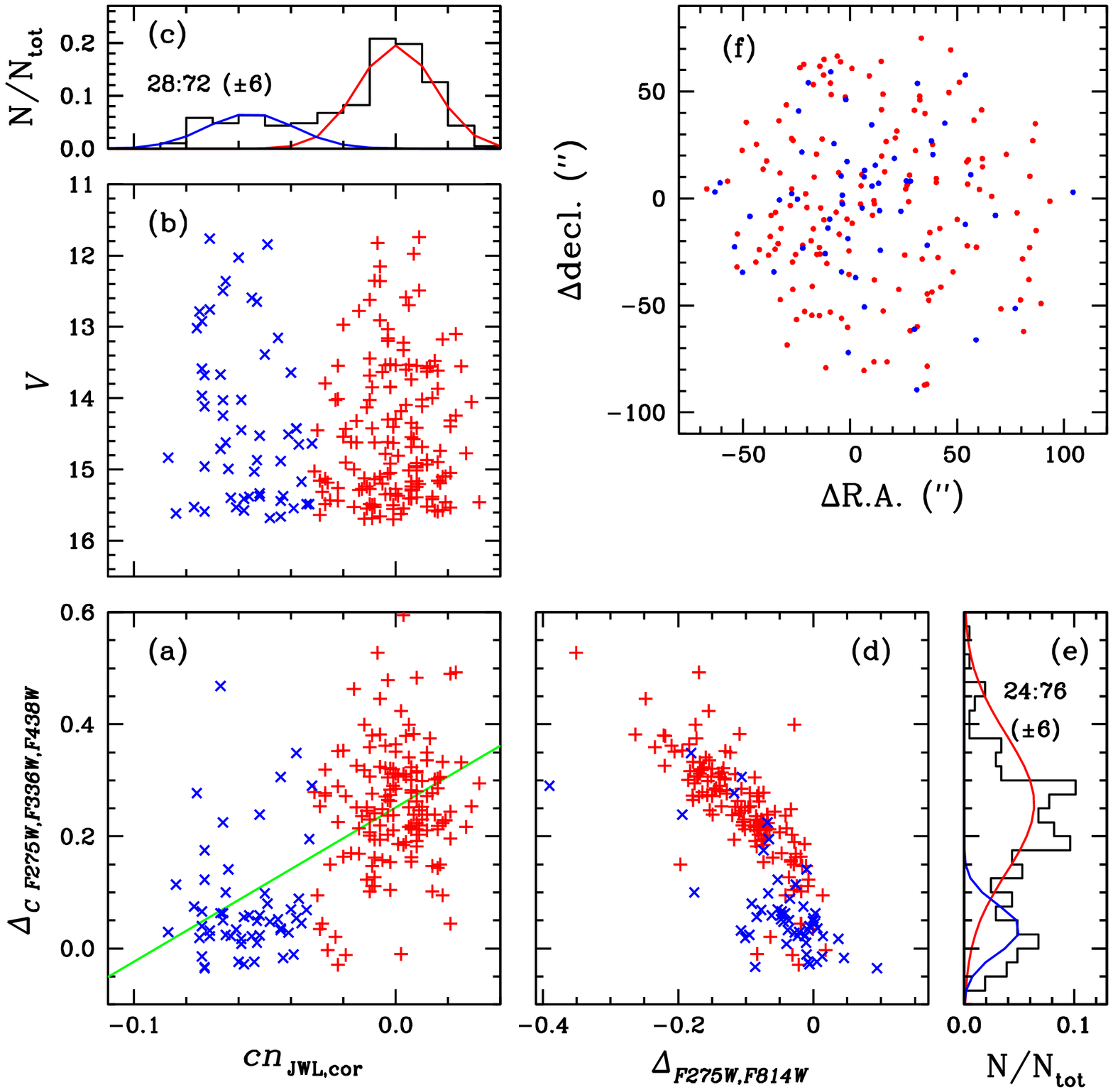}
\caption{(a) Comparison of our \cnjwl\ with the \dtrio\ indices \citep{piotto15,milone17}.
The blue crosses denote the \cnw\ and the red plus signs denote the \cns\ RGB stars from our EM estimator.
The green solid line represents the linear fit to the data.
In general, our result is in good agreement with that of \citet{piotto15}.
(b) \vcn\ CMD. 
(c) Histogram of the \cnjwlcor\ distribution. The populational number ratio from our \cnjwlcor\ for the common RGB stars is \nrgb\ = 28:72 ($\pm$6).
(d) A plot of \dtrio\ versus \dc\ \citep[also known as a chromosome map; see][]{milone17}.
(e) Histogram of the \dtrio\ distribution with a number ratio of 24:75 ($\pm$6). The frequency of the first generation of stars is in good agreement with that of \citet{milone17}, 0.294 $\pm$ 0.023.
(f) Positions of the RGB stars used in our analysis.}
\label{fig:hst}
\end{figure}

\subsubsection{A comparison with \citet{piotto15}}
We also made a comparison of our photometry with the \hst\  $UV$ photometry by \citet{piotto15} and we show our results in Figure~\ref{fig:hst}.
We obtained the RGB number ratios of \nrgb\ = 28:72 ($\pm$6) and 24:76 ($\pm$6) for our \cnjwlcor\ and \dtrio,\footnote{See equations (1) and (2) of \citet{milone17} for the definitions of the \dc\ and the \dtrio\ indices.} respectively.
Our result is in excellent agreement with that by \citet{milone17}, who obtained the fraction of FG stars of 0.294 $\pm$ 0.023 (see their Table~2).
It should be emphasized that a nice correlation can be seen between our \cnjwlcor\ and  \dtrio.
As shown in Figure~2 of \citet{lee17}, the \hst\ F336W filter can cover the NH molecular band absorption features as a whole.
Also, unlike other filter systems being used in ground-based observations, it is free from the atmospheric extinction, leading the \dtrio\ to be sensitive to the NH abundances.
Therefore, it is not a surprise that there is a tight correlation between our \cnjwlcor\ and \dtrio, which is a photometric analog of the positive CN-NH correlation seen in GC RGB stars.
As we already pointed out, since the \dtrio\ measures OH, NH, CN, CH, C$_2$, Mg, and Ca at the same time, and absorption features from these elements experience different degrees of the luminosity effects, the \dtrio\ may suffer from some confusion.
We conclude that our \cnjwlcor\ index is as good as the \hst\ \dtrio\
and it provides a very powerful means to investigate the MSPs of GCs.

\begin{deluxetable}{crr}
\tablecaption{RGB bump $V$ magnitudes
\label{tab:bump}}
\tablenum{4}
\tablewidth{0pc}
\tablehead{
\multicolumn{1}{c}{} &
\multicolumn{1}{c}{\cnw} &
\multicolumn{1}{c}{\cns}
}
\startdata
\vbump                 & 13.667  $\pm$ 0.030 & 13.631 $\pm$ 0.030\\
\enddata
\end{deluxetable}

\begin{figure}[b]
\epsscale{1.0}
\figurenum{10}
\plotone{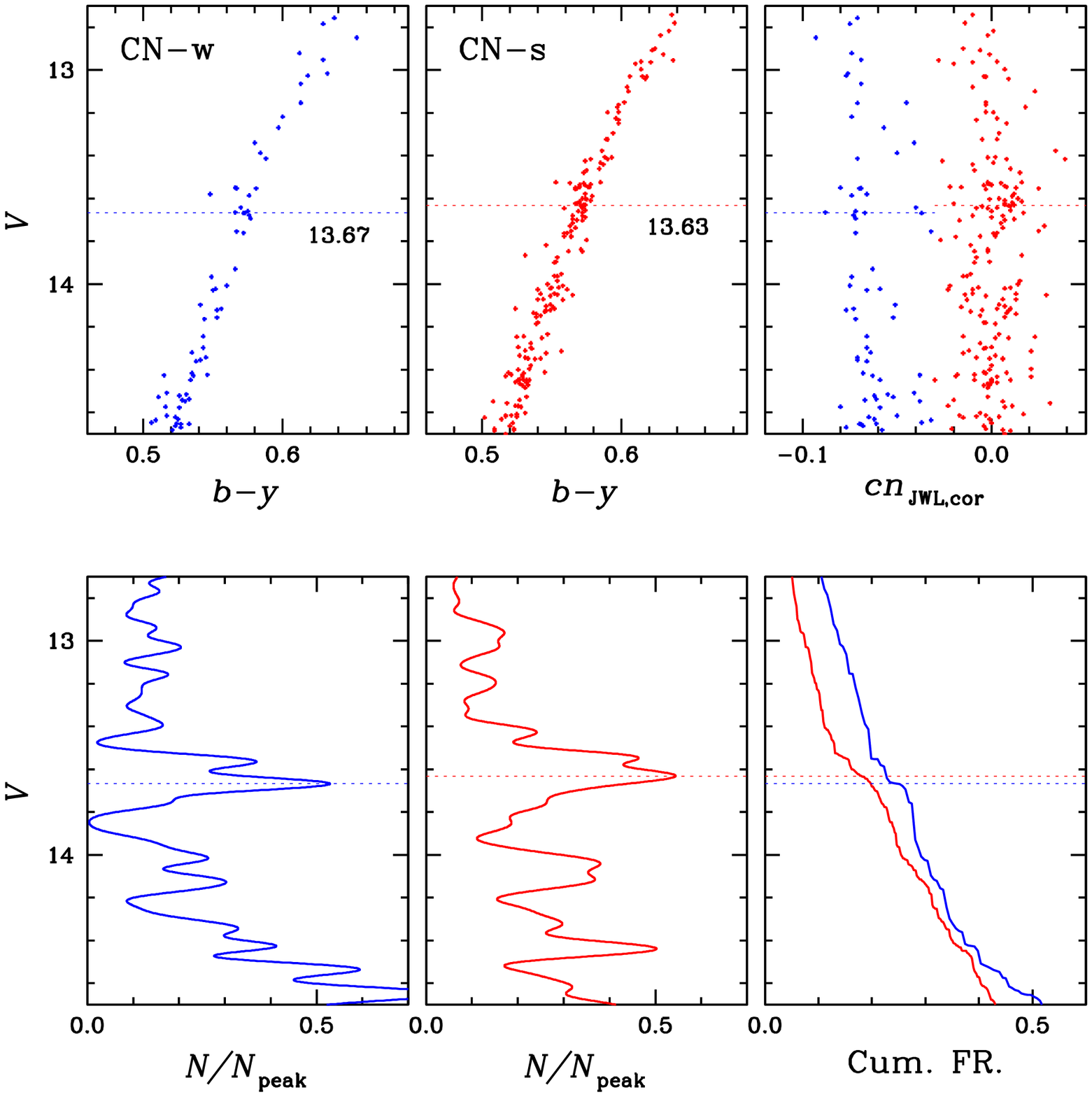}
\caption{(Upper panels) Plots of \vby\ and \vcn\ CMDs of the \cnw\ and the \cns\ RGB stars in NGC~6752.
The dashed lines are the $V$ magnitude of the RGB bump, \vbump.
We obtained \vbump\ = 13.67 mag for the \cnw\ population and 13.63 mag for the \cns\ population.
(Lower panels) Generalized differential and cumulative LFs for each population.
}\label{fig:bump}
\end{figure}

\subsubsection{RGB bump magnitude: \vbump}
Despite helium being the second most abundant element, the helium abundance of the GC RGB stars is difficult to measure.
Direct spectroscopic measurements of the helium abundance of the cool giant stars in NGC~2808 and NGC~5139 have been conducted using the chromospheric \ion{He}{1} absorption line at $\lambda$10830 \AA\ \citep[e.g., see][]{pasquini11,dupree13}, which is sensitively dependent on the structure and dynamics of the atmospheric models.
The helium abundance of the blue HB (BHB) stars cooler than the Grundahl jump\footnote{See \citet{grundahl99} for the definition of the Grundahl jump.} can also be measured using the photospheric \ion{He}{1} absorption line at $\lambda$5876 \AA \citep[e.g., see][]{villanova09,marino14}.
In fact, \citet{villanova09} measured the helium abundance for the five O-rich, Na-poor NGC~6752 BHB stars, $Y$ = 0.245 $\pm$ 0.012.
Note that these BHB stars are equivalent to the \cnw\ population in our study.
Unfortunately, due to the low temperature of the O-poor, Na-rich BHB stars, which is equivalent to the \cns\ population, \citet{villanova09} was not able to measure the helium abundance.
Consequently, they was not able to see if there exists a difference in the helium abundance between the two BHB populations in NGC~6752.

Alternatively, indirect methods making use of the helium sensitive characteristics, such as the RGB bump luminosity \citep[e.g.][]{cassisi13, largioia18, lee15, lee17, milone15} and the number ratio of HB to RGB stars \citep{buzzoni83}, have been frequently used to estimate the helium abundances.

In order to investigate the relative helium abundance difference between the two populations, we compared the RGB bump $V$ magnitudes using our \cnjwlcor\ CMD.
In Figure~\ref{fig:bump}, we show CMDs around the RGB bumps and the differential and the cumulative luminosity functions (LFs), finding \vbump\ = 13.667 $\pm$ 0.030 for the \cnw\ and 13.630 $\pm$ 0.030 for the \cns\ RGB stars (see also Table~\ref{tab:bump}).
Following the same method that we employed in our previous studies \citep[][and references therein]{lee15,lee17}, we obtained the difference in the helium abundance of \dy\ = 0.016 $\pm$ 0.016, in the sense that the mean helium abundance of the \cns\ RGB stars is slightly more enhanced.
During our calculation, we assumed no metallicity and age differences between the two populations.

We emphasize that our relative difference in the helium abundance between the two populations is in excellent agreement with those by \citet{milone13} and \citet{largioia18}, who obtained \dy\ $\approx$ 0.015 and 0.018 $\pm$ 0.004, respectively.

Based on our \cnjwlcor\ index, we conclude that the two RGB populations in NGC~6752 exhibit very different chemical compositions: 
the \cnw\ population has high [O/Fe] and low [Na/Fe], [N/Fe] abundances, while the \cns\ population has low [O/Fe] and high [Na/Fe], [N/Fe] abundances with the sign of a slight helium enhancement.
Also, the evolution of nitrogen abundance is discrete while those of oxygen and sodium are continuous between the two populations.
Hence, in the context of the self-enrichment scenario, the \cnw\ population is equivalent to the first generation (FG), while the \cns\ population is the second generation (SG) of the stars.

\begin{deluxetable*}{ccrrcrrcrr}[t]
\tablecaption{Centers of NGC~6752
\label{tab:cnt}}
\tablenum{5}
\tablewidth{0pc}
\tablehead{
\multicolumn{1}{c}{} &
\multicolumn{1}{c}{} &
\multicolumn{2}{c}{Simple Mean} &
\multicolumn{1}{c}{} &
\multicolumn{2}{c}{Half-sphere} &
\multicolumn{1}{c}{} &
\multicolumn{2}{c}{Pie-slice }\\
\cline{3-4}\cline{6-7}\cline{9-10}
\colhead{} & \colhead{} & 
\colhead{$\Delta\alpha$(\arcsec) } & \colhead{$\Delta\delta$(\arcsec)} &
\colhead{} &
\colhead{$\Delta\alpha$(\arcsec) } & \colhead{$\Delta\delta$(\arcsec)} &
\colhead{} &
\colhead{$\Delta\alpha$(\arcsec) } & \colhead{$\Delta\delta$(\arcsec)} }
\startdata
All    & &     1.2 &  1.4 &&   $-$0.7 &    0.1 && $-$2.4 & $-$1.0 \\
\cnw\  & &  $-$1.1 &  0.4 &&   $-$1.5 & $-$2.0 && $-$2.0 &    3.1 \\
\cns\  & &     1.9 &  1.8 &&      2.9 &    0.3 && $-$2.3 & $-$1.2 \\
\enddata
\end{deluxetable*}

\subsubsection{Centers}
For the first step in our investigation of the structural properties of individual populations, we calculated the centers of the population using three different methods; the arithmetic mean, the half-sphere and the pie-wedge methods \citep[see][for detailed discussions]{lee15,lee17}.

Our results are shown in Table~\ref{tab:cnt}.
Depending on the calculation methods, the coordinates of the center of each population can be slightly different. 
However, the angular separations between the centers of the two populations are negligibly small compared to the core or the half-light radii of NGC~6752, 10\arcsec\ and 115\arcsec, respectively \citep{harris96}.
Therefore, we conclude that the coordinates of the centers of individual populations are almost identical.

\begin{figure}
\epsscale{1.0}
\figurenum{11}
\plotone{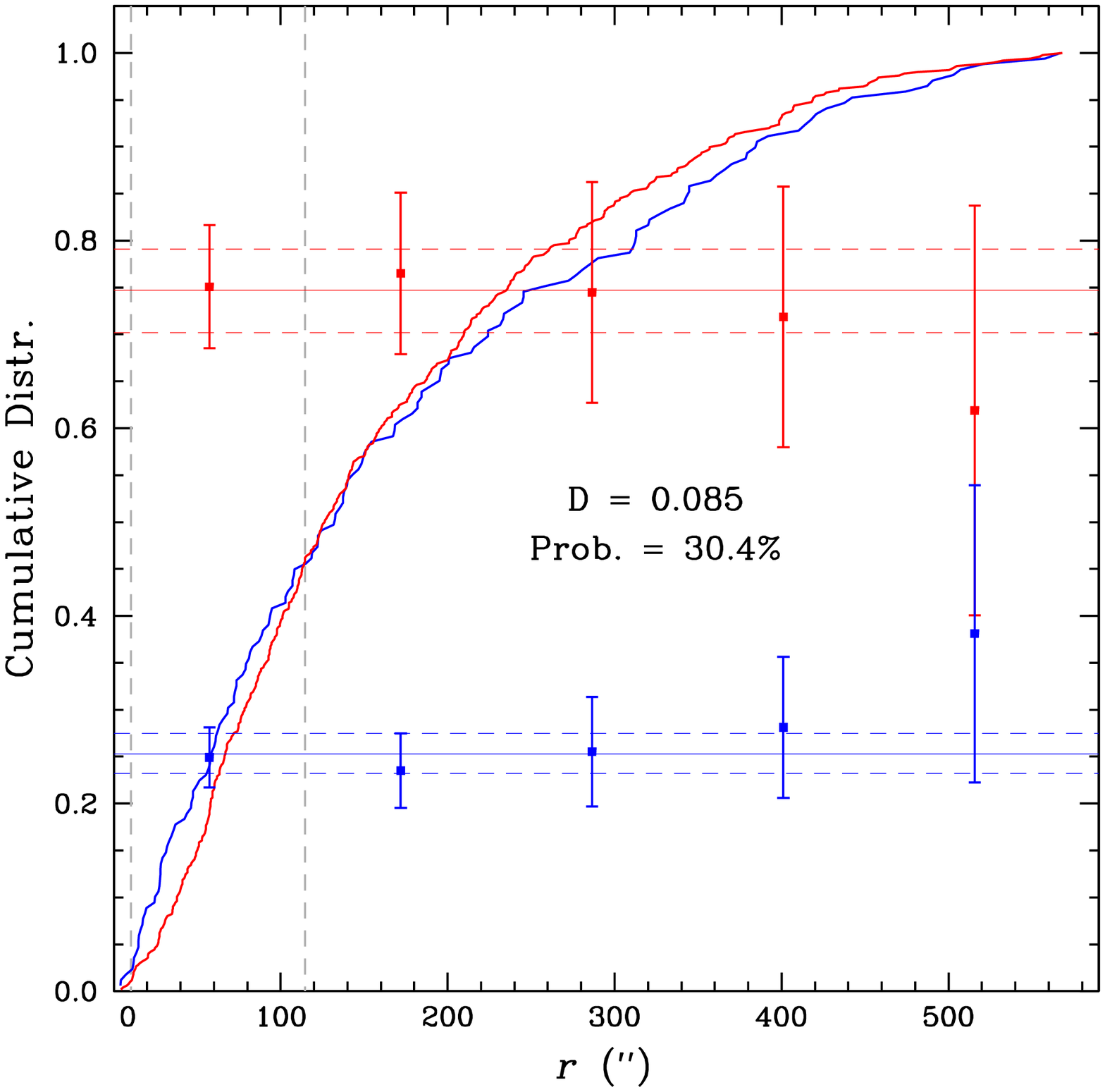}
\caption{Cumulative radial distributions of the \cnw\ (blue) and the \cns\ (red) RGB populations in NGC~6752. 
A K--S test shows that both populations are most likely drawn from an identical parent distribution.
The blue and the red horizontal lines denote the fractions of the \cnw\ and the \cns\ populations, n(\cnw)/n(\cnw\ + \cns) = 0.253 $\pm$ 0.022 and n(\cns)/n(\cnw\ + \cns) = 0.747 $\pm$ 0.044.
}\label{fig:rad}
\end{figure}

\begin{figure}
\epsscale{1}
\figurenum{12}
\plotone{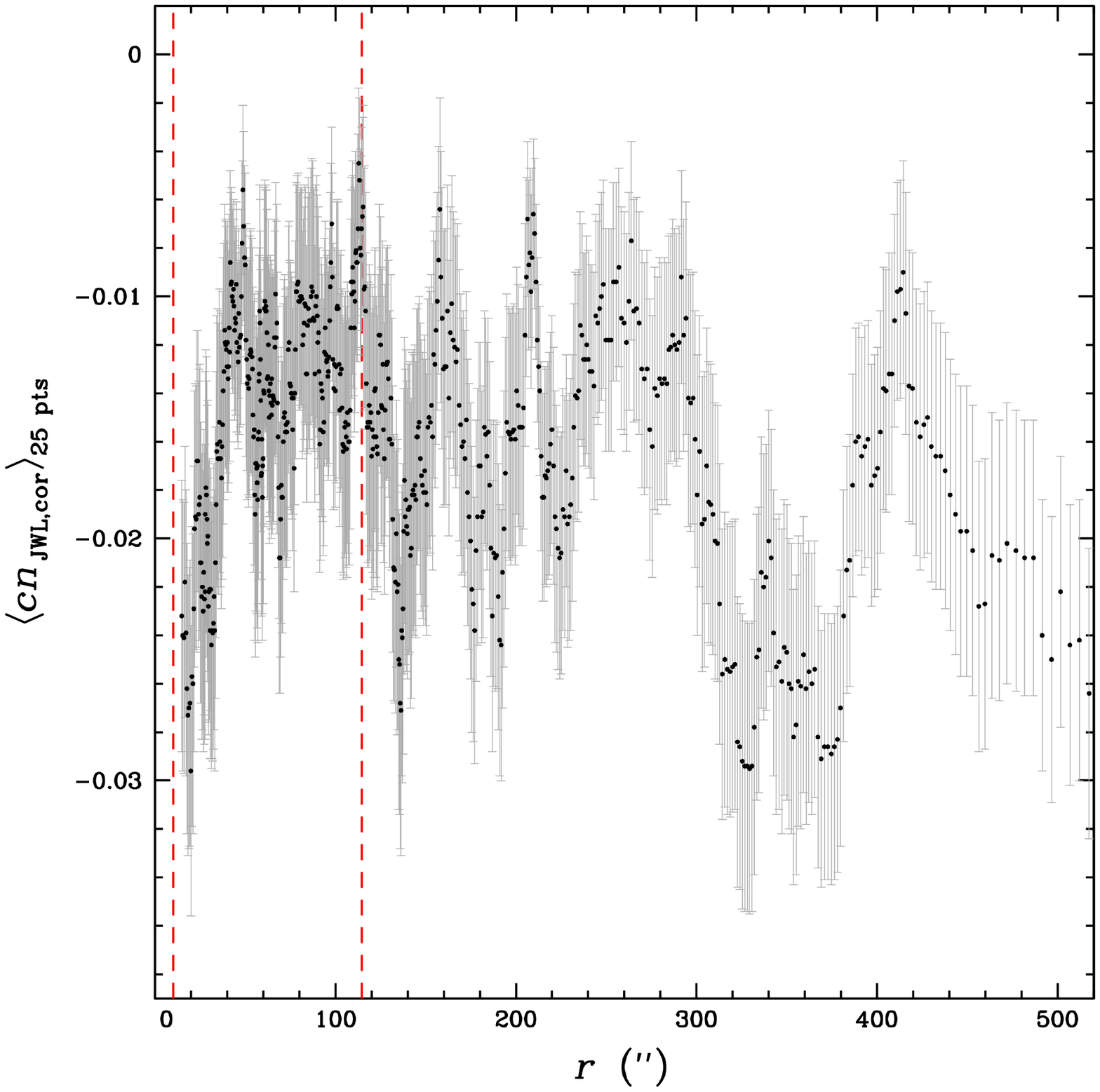}
\caption{Moving average of the adjacent 25 points for our \cnjwlcor\ index, \mavg, of the RGB stars against the radial distance from the center.
The thin gray error bars represent the standard error of the mean and the vertical red dashed lines denote the core and the half-light  radii of the cluster.
The \mavg\ remains flat and does show any significance variation against the radial distance, which is consistent with Figure~\ref{fig:rad}.
}\label{fig:runavg}
\end{figure}

\subsubsection{cumulative radial distributions}
The cumulative radial distribution has been widely used to compare the structural properties between the MSPs in GCs, although the time scale for the homogenization of the radial distributions of the MSPs appears to be somewhat uncertain \citep[e.g., see][and references therein]{lee15,lee17}.

In Figure~\ref{fig:rad}, we show the cumulative radial distribution of the \cnw\ and the \cns\ RGB stars in NGC~6752.
We performed the Kolmogorov--Smirnov (K--S) test, obtaining the significance level for the null hypothesis that the both distributions are drawn from the same distribution of $p$ = 0.304, with a K--S discrepancy of 0.085.
Our result suggests that the radial distributions of both populations are almost identical, up to more than 4.5 half-light radii.
In the figure, we also show the fractions of the \cnw\ and the \cns\ RGB populations, n(\cnw)/n(\cnw\ + \cns) = 0.253 $\pm$ 0.022 and n(\cns)/n(\cnw\ + \cns) = 0.747 $\pm$ 0.044, respectively.
Owing to the small number statistics, there exists some mild fluctuation at large distances from  the center, $r \geq 250\arcsec$ ($\approx$ 2$r_h$), but the fractions of each population are flat and do not show any significant variations against the radial distance. 

The moving average of the \cnjwlcor\ index can provide a very useful diagnostic for examining the radial variation of the population ratios.
In Figure~\ref{fig:runavg}, we show the moving average of the adjacent 25 points for the \cnjwl\ index, \mavg, against the radial distance.
The moving average shows some small-scale local fluctuations at insignificant levels and any large scale gradient cannot be seen in NGC~6752, i.e., the absence of the radial CN gradient.
It should be reminded that this was also found in M5 \citep{lee17}.
According to \citet{vesperini13}, the similar radial distributions for the MSPs in NGC~6752, with flat number ratios up to more than 4.5$r_h$ imply that NGC~6752 has already achieved the complete mixing.

\begin{figure}
\epsscale{1.0}
\figurenum{13}
\plotone{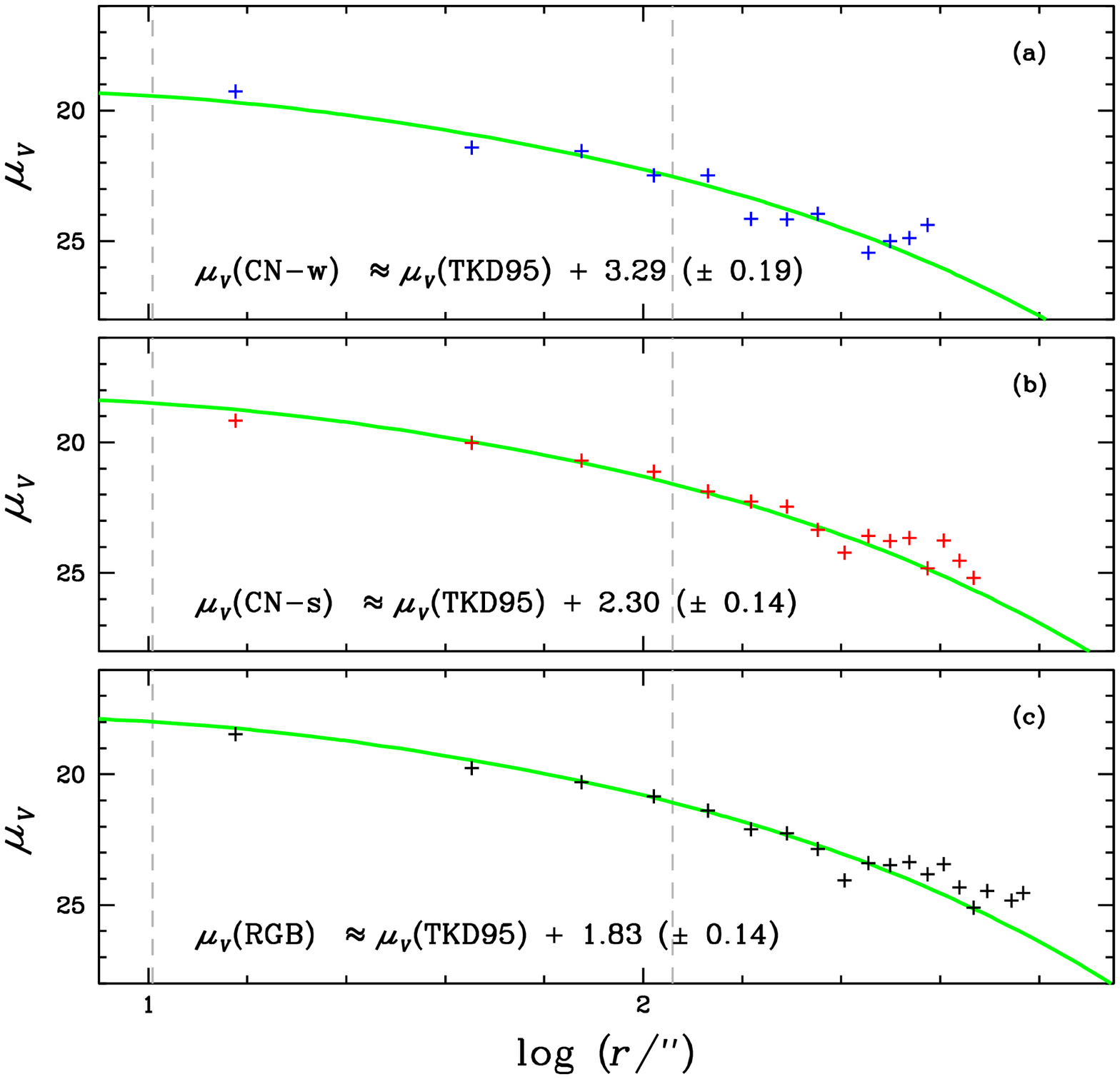}
\caption{Surface brightness profiles for the  RGB stars with 
$-$2 $\leq$ \vvhb\ $\leq$ 2 mag. (a) \cnw\ RGB stars only; 
(b) \cns\ stars only; and (c) all RGB stars in NGC~6752.
The Green lines denote the Chebyshev polynomial fit of the surface brightness
profile of NGC~6752 by \citet{trager95}.
}\label{fig:sbp}
\end{figure}

\subsubsection{Surface brightness profiles}
In our previous study \citep{lee15,lee17}, we showed that the surface brightness profile (SBP) of GC can provide useful information on the structural property of the MSPs in GCs.
Using the same method employed in our previous study, we derived the SBPs of NGC~6752 RGB stars in both populations.
The SBPs of the bright RGB stars in NGC~6752 are shown in Figure~\ref{fig:sbp}, along with the Chebyshev polynomial fit of the cluster by \citet{trager95}.
The figure clearly shows that our SBP measurements for each population are in excellent agreement with that of \citet{trager95}, up to 8\arcmin\ (more than 4$r_h$) from the center, confirming our result that both populations have an identical radial distribution.
These similar SBPs for both the \cnw\ and the \cns\ populations also imply that they are already reached in a dynamically homogenized state.

\begin{figure}
\epsscale{1.0}
\figurenum{14}
\plotone{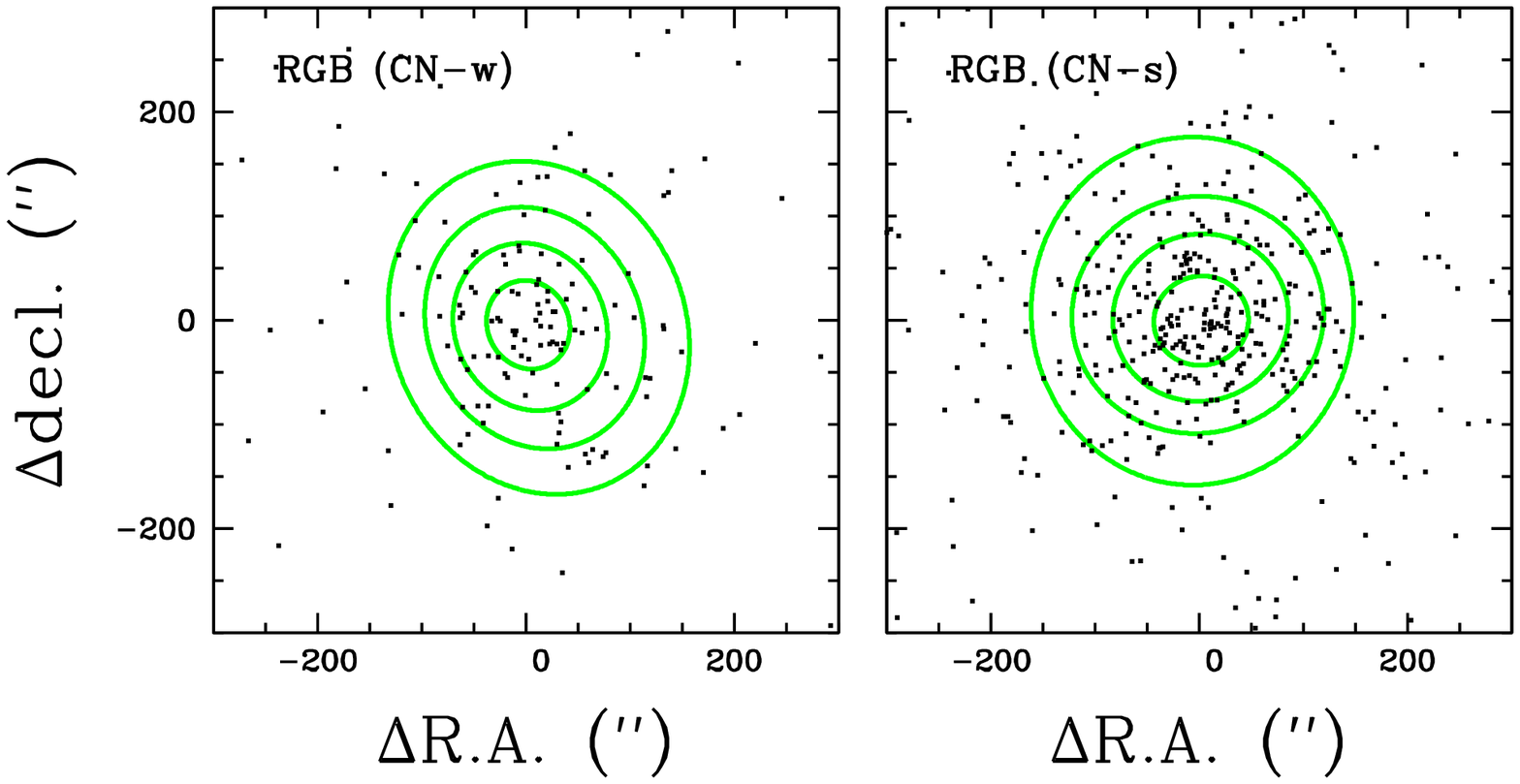}
\caption{Spatial distributions of the \cnw\ and the \cns\ RGB stars in NGC~6752.
The solid green lines denote the isodensity contours.
Note that the spatial distribution of the \cnw\ RGB population 
is more elongated along the NNW--SSE direction.
}\label{fig:density}
\end{figure}

\begin{deluxetable}{lrrrr}
\tablecaption{Ellipse fitting parameters for NGC~6752 RGB stars
\label{tab:ellipse}}
\tablenum{6}
\tablewidth{0pc}
\tablehead{
\multicolumn{1}{c}{} &
\multicolumn{1}{c}{grid} &
\multicolumn{1}{c}{$\theta$} &
\multicolumn{1}{c}{$b/a$ } &
\multicolumn{1}{c}{$e$}}
\startdata
\cnw    & 0.9 &  117.2 $\pm$  3.4 &  0.902 $\pm$  0.038 &  0.098 \\
        & 0.7 &  115.7 $\pm$  2.9 &  0.879 $\pm$  0.032 &  0.121 \\
        & 0.5 &  116.0 $\pm$  2.3 &  0.857 $\pm$  0.025 &  0.143 \\
        & 0.3 &  114.7 $\pm$  1.6 &  0.857 $\pm$  0.018 &  0.143 \\
        & & & & \\ 
\cns    & 0.9 &   10.7 $\pm$  3.2 &  0.940 $\pm$  0.037 &  0.060 \\
        & 0.7 &   15.1 $\pm$  2.7 &  0.947 $\pm$  0.032 &  0.053 \\
        & 0.5 &    9.9 $\pm$  2.2 &  0.939 $\pm$  0.025 &  0.061 \\
        & 0.3 &   90.4 $\pm$  1.6 &  0.927 $\pm$  0.018 &  0.073 \\
\enddata
\end{deluxetable}

\begin{figure}
\epsscale{1.0}
\figurenum{15}
\plotone{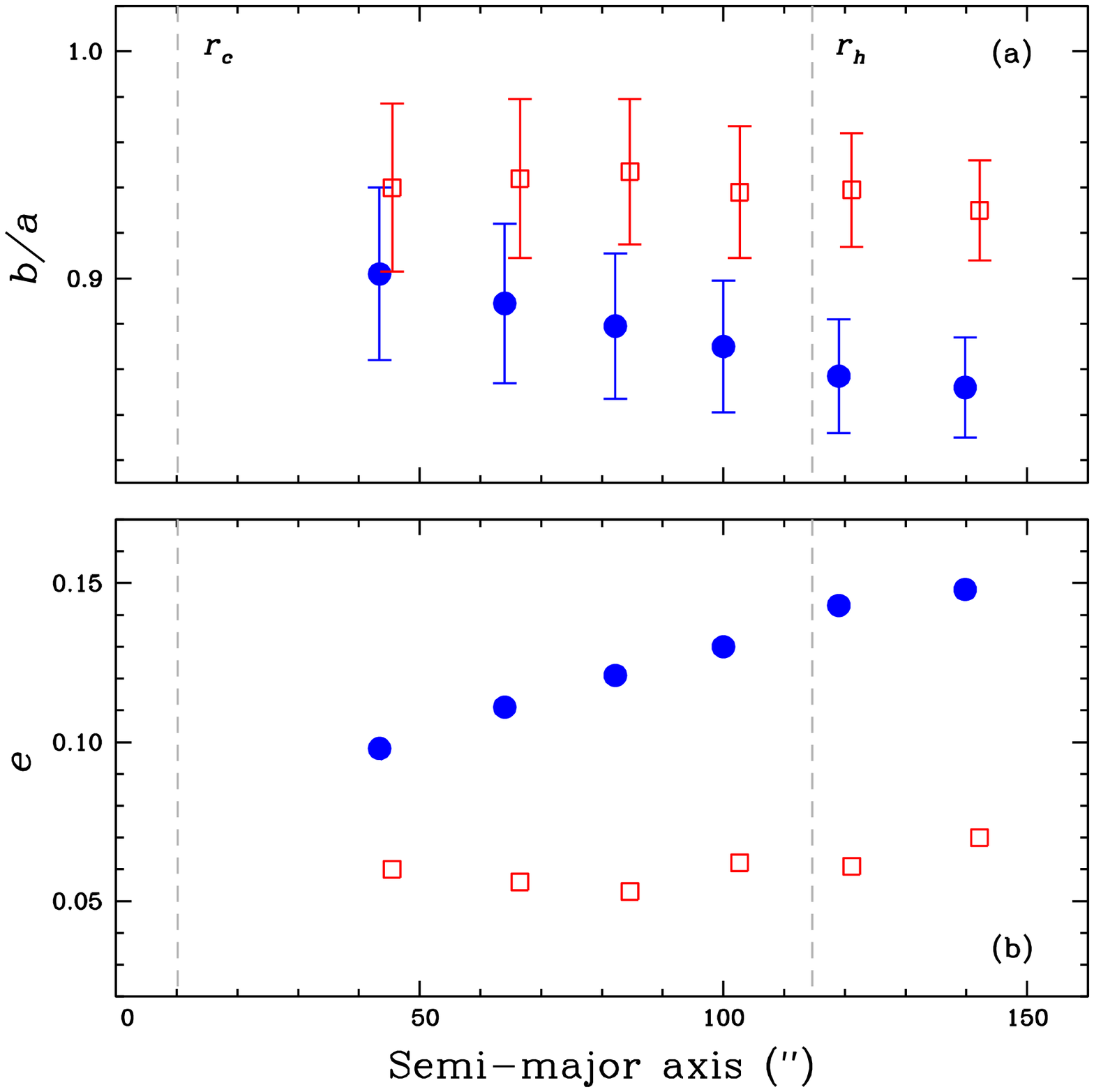}
\caption{Axial ratio, $b/a$, and the ellipticity, $e$ ($= 1 - b/a$), of the \cnw\ (blue) and the \cns\ (red) populations against the major axis, $a$.
}\label{fig:ellip}
\end{figure}

\subsubsection{Spatial distributions}
Our previous study of M5 revealed that the two RGB populations have very similar radial distributions, with flat number ratios but very different spatial distributions, which are closely linked with the different internal kinematics between the two stellar populations in M5 \citep{lee17}. 
Here, we explore the spatial distributions of RGB stars in NGC~6752.
Figure~\ref{fig:density} shows the projected spatial distributions for the two RGB populations in NGC~6752.
As can be seen in the figure, the projected distribution of the \cnw\ population is more spatially elongated along the NNE--SSW direction while that of the \cns\ population is not.
In Table~\ref{tab:ellipse}, we show the position angle, the axial ratio, $b/a$, and the ellipticity, $e$ $(= 1 - b/a)$. 
The radial distributions of the axial ratio and the ellipticity are shown in Figure~\ref{fig:ellip}.
The ellipticity of the \cnw\ population is much larger than that of the \cns\ population.
The difference in the spatial distributions between the two RGB populations cannot be inferred from the cumulative radial distributions or the SBPs of each population.
As shown below, it is strongly believed that the elongated projected spatial distribution of the \cnw\ RGB population is closely linked with its faster projected rotation, as we already showed for M22 and M5 \citep{lee15,lee17}.
Our result is another vivid example of the importance of studying the spatial distributions of MSPs in GCs at rather large radial distances to examine the structural differences.

\begin{figure}
\epsscale{1.0}
\figurenum{16}
\plotone{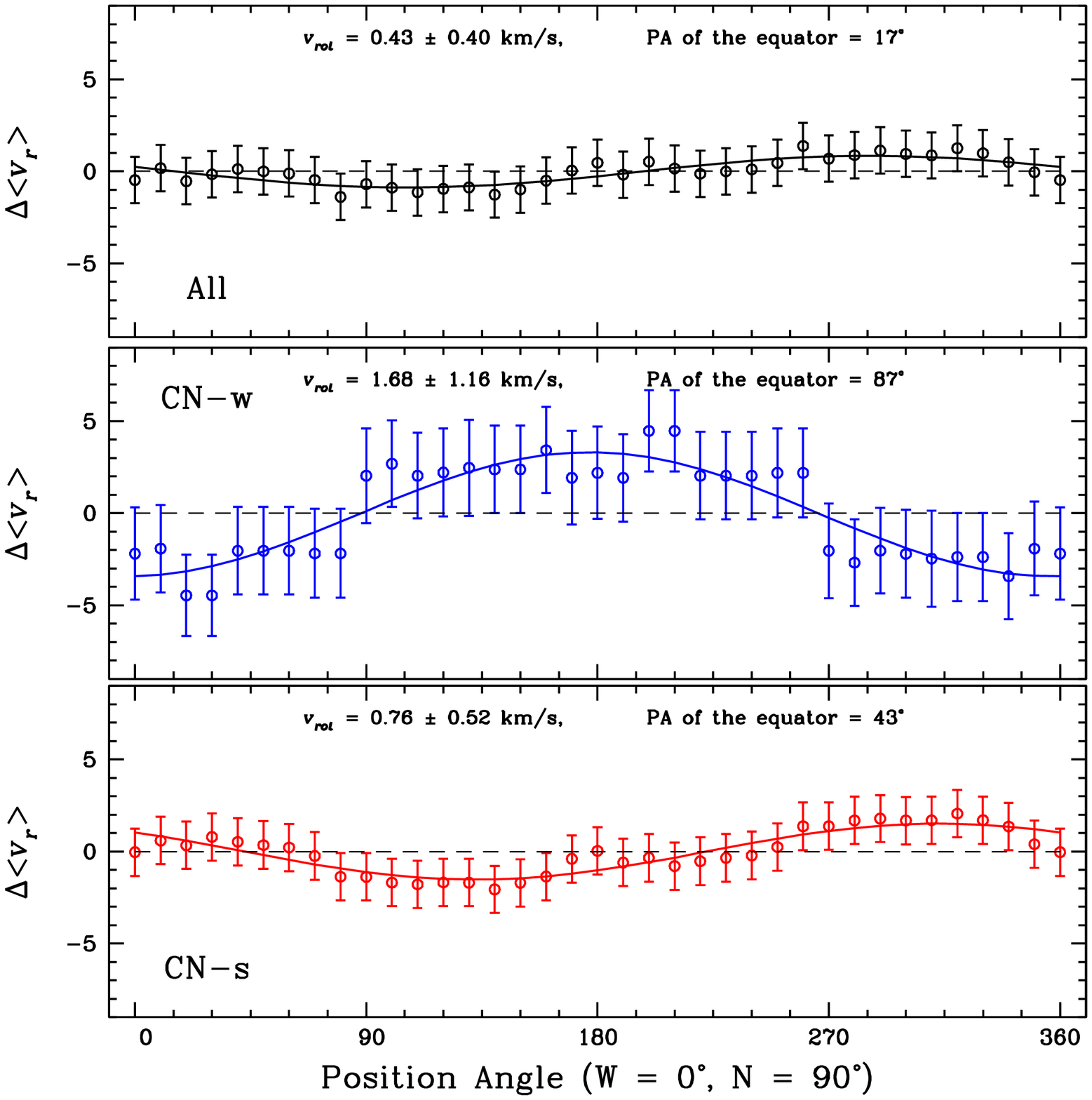}
\caption{Difference in the mean radial velocities between both 
hemispheres against their position angles, where the net rotation is the half
of the amplitude of the sinusoidal fit.
It is evident that the \cnw\ population shows a substantial rotation. 
}\label{fig:rot}
\end{figure}

\subsubsection{Projected Rotations}
A few previous studies of the the projected rotation of NGC~6752 are available.
For example, \citet{lane10} reported that NGC~6752 does not show any signs of rotation.
On the other hand, using the same data, \citet{lardo15} obtained a non-negligible rotation with \arot\ $\approx$ 0.7 \kms.
Note that \arot\ of \citet{lardo15} is twice as large as the rotational velocity defined by \citet{lane10} and our previous works for M22 and M5 \citep[see also Appendix A of][]{bellazzini12}.
Regardless of the real value of the amplitude of the projected rotation of the cluster, what was neglected in the previous studies is the rotational properties of the individual populations in NGC~6752.
We explore the projected rotation of RGB stars in each population.

We derived the amplitude and the position angle of the projected mean rotation of the cluster using the method that we employed in our previous study \citep{lee15,lee17}.
The basic idea is that, assuming an isothermal sphere, the mean rotation velocity can be estimated by dividing a GC in half at a given position angle and by calculating differences between the average radial velocities in the two hemispheres. 
Then, the net rotation of a GC is half of the amplitude of the sinusoidal fit.

For our calculations, we used the radial velocity measurements by \citet{carretta07}. 
A comparison of the radial velocities for 84 membership stars by \citet{lane10} and \citet{carretta07} gives $\Delta v_r$ = 0.36 $\pm$ 1.03 \kms, in the sense that 
the mean radial velocity from \citet{lane10} is slightly larger than that from \citet{carretta07}. 
What concerns us is rather large value of the standard deviation of the difference in the radial velocities between the two measurements, because a small-scale structure can be eliminated if the internal error is sufficiently large.
We are not able to assess the quality of the radial velocity measurements by \citet{carretta07} and \citet{lane10} and we chose to use the measurements by \citet{carretta07} for two reasons:
(i) the spectral resolution from \citet{carretta07} is higher than that from \citet{lane10} ($\lambda/\Delta\lambda$ $\approx$ 23,000 versus 10,000), and 
(ii) the signal-to-noise ratios of individual spectra from \citet{carretta07} are expected to be larger than those from \citet{lane10}.

We show the differences in the mean radial velocities between both hemispheres as a function of the position angle and the best-fitting sine function in Figure~\ref{fig:rot}.
For all RGB stars, we obtained a negligibly small mean rotation velocity, $v_{\rm rot}$ = 0.4 $\pm$ 0.4 \kms\, and a position angle of the equator of 17$^\circ$. 
The rotation velocities of the individual populations are given in the middle and bottom panels of Figure~\ref{fig:rot}.
We obtained an average projected rotation of 1.7 $\pm$ 1.2 \kms\ with a position angle of 87\degr\ for the \cnw\ population, and for the \cns\ population we obtained an average projected rotation of 0.8 $\pm$ 0.5 \kms, with a position angle of 43\degr.
It can be clearly seen that the \cnw\ population exhibits a substantial net projected rotation.
Moreover, if the fast rotation of the the \cnw\ population is responsible for its elongated distribution, it can be said that the position angle of the equator of the projected rotation of the \cnw\ population (i.e., along the N--S direction) is marginally in agreement with the projected spatial distribution of the \cnw\ population (NNE--SSW) as shown in Figure~\ref{fig:density} and Table~\ref{tab:ellipse}.

As we mentioned above, our previous study of M5 revealed for the first time that the \cns\ population, which is thought to be the SG, has a substantial rotation, \vrot\ = 2.1 $\pm$ 0.9 \kms, while the \cnw\ population has no net rotation, leading us to propose that the different rotational properties are qualitatively in accord with the normal GC formation scenario proposed by \citet{bekki10}.
Interestingly, NGC~6752 is the opposite:
The FG of the stars appears to rotate faster than the SG of the stars in NGC~6752.
The \cnw\ population in NGC~6752 is thought to be a good example of showing a structure-kinematics coupling.
The cumulative radial distribution suggests that the two populations in NGC~6752 are in the state of the relaxed system, but the projected spatial distribution and rotation imply that the two populations are yet to be homogenized.
In the future, more systematic and comprehensive observations for radial velocity measurements of NGC~6752 RGB stars will be very desirable and will shed more light on the kinematic property of the cluster.
At the same time, more robust and through theoretical calculations to reveal the dynamical evolution of the GC systems are definitely needed.

\begin{figure}
\epsscale{1.0}
\figurenum{17}
\plotone{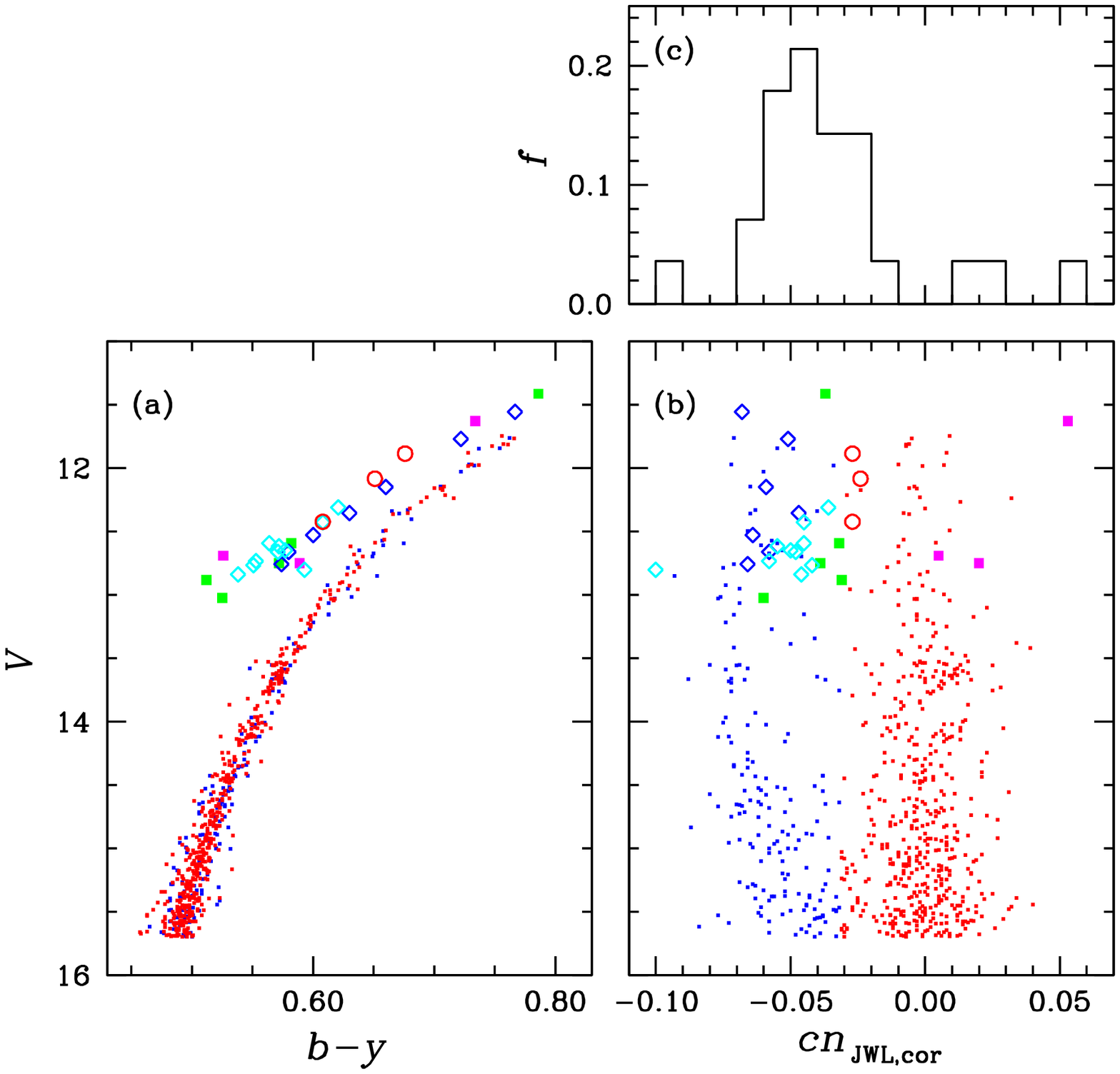}
\caption{(a) A $(b-y)$ versus $V$ CMD of RGB stars with $-2 \leq$ \vvhb\ $\leq$ 2 mag and all AGB stars detected in this study. 
The blue and red dots denote the \cnw\ and the \cns\ RGB stars, while the blue open diamonds and the red open circles denote the \cnw\ and the \cns\ AGB stars studied by \citet{campbell13} and \citet{lapenna16}. 
The cyan open diamonds are the \cnw\ population from our photometry, but with the intermediate population based on the sodium abundance measurements by \citet{lapenna16}.
The filled green and magenta squares are for the \cnw\ and \cns\ AGB stars without any previous spectroscopic study.
(b) Same as (a) but for \vcn\ CMD. 
As shown, NGC~6752 appears to have some extreme \cns\ AGB stars shown with magenta squares.
(c) A histogram for the AGB stars.
}\label{fig:agbcmd}
\end{figure}

\subsection{Asymptotic giant stars}\label{s:agb}
The effective temperatures of GC AGB stars are not hot enough to completely suppress the CN molecular band formations in their atmospheres.
Therefore, if the GC AGB stars exhibit a variation in the CN abundance, our \cnjwlcor\ index will provide a powerful means to explore the AGB populations in GCs, especially in the central parts of GCs.

It is well known that the study of AGB stars has some fundamental limitations.
First, AGB stars are much rarer than their progenitors  (i.e., MS, RGB, and HB stars) and the statistical fluctuation that arises from the small number statistics cannot be avoided.
For example, the stochastic truncation in the outer part of the cluster is most likely responsible for the lack of the \cns\ AGB population in M5 \citep{lee17}.
Second, due to the diversity of the post-HB evolutionary passages of the low-mass stars on the CMD, every HB star does not pass through the AGB sequence and some failed to evolve into the AGB (the so-called \agbm\ star), which hinder us from performing the comprehensive census of the AGB populations.

\begin{figure}
\epsscale{1.0}
\figurenum{18}
\plotone{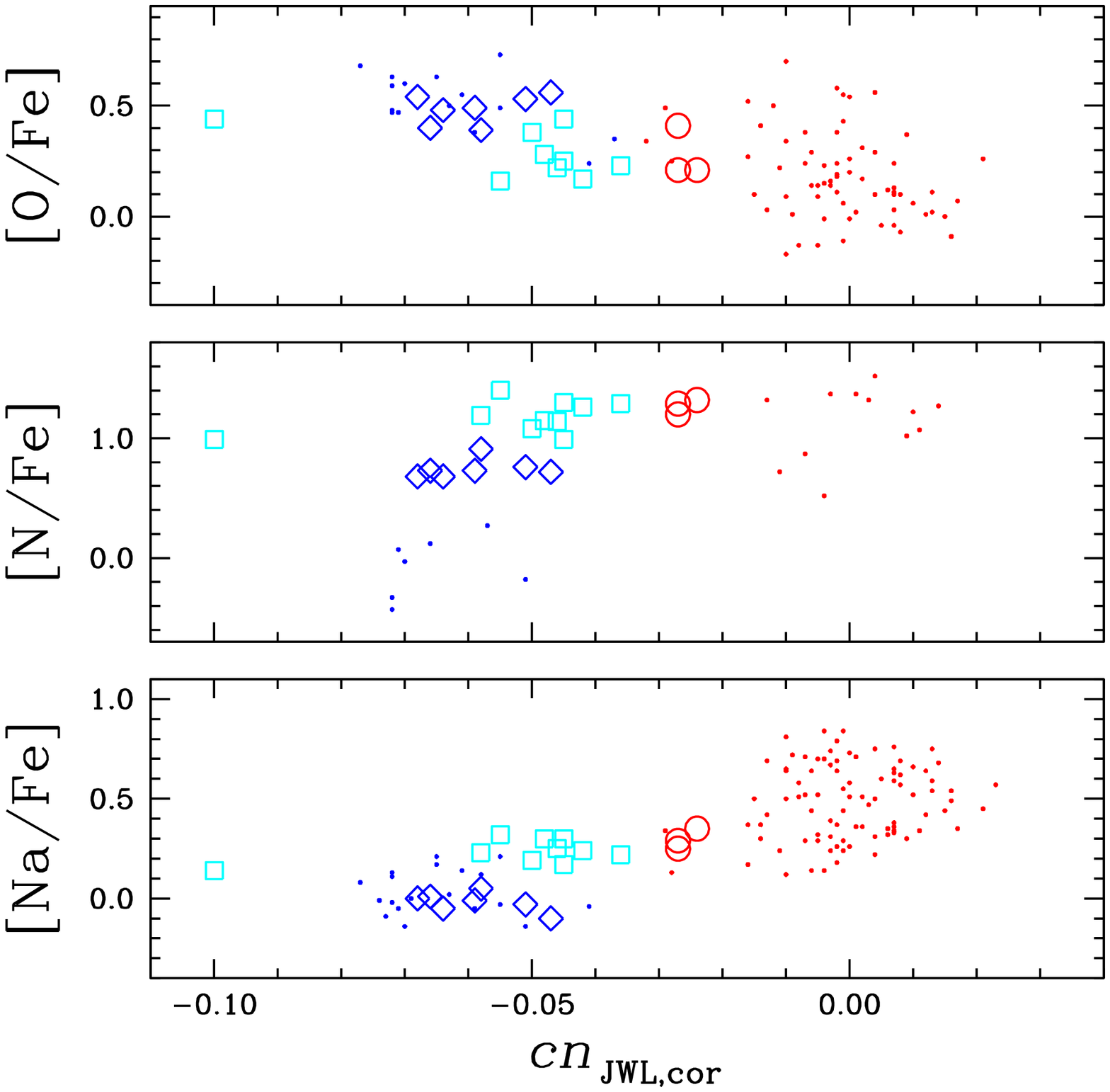}
\caption{Elemental abundances against \cnjwlcor.
The blue and red dots denote the \cnw\ and the \cns\ RGB stars, while the blue open diamonds and the red open circles denote the \cnw\ and the \cns\ AGB stars studied by \citet{lapenna16}.
The cyan open squares are the \cnw\ population from our photometry, but with the intermediate population based on the sodium abundance measurements from \citet{lapenna16}.
For the RGB stars, [O/Fe] and [Na/Fe] are from \citet{carretta07}, while [N/Fe] is from \citet{yong08}.
Note that the nitrogen abundance from the CN molecular bands by \citet{lapenna16} is higher than that from the NH molecular bands by \citet{yong08}.
The nine AGB stars classified as an intermediate population by \citet{lapenna16} are most likely the \cnw\ population from our photometric point of view.
}\label{fig:agbspec}
\end{figure}

\begin{figure}
\epsscale{1}
\figurenum{19}
\plotone{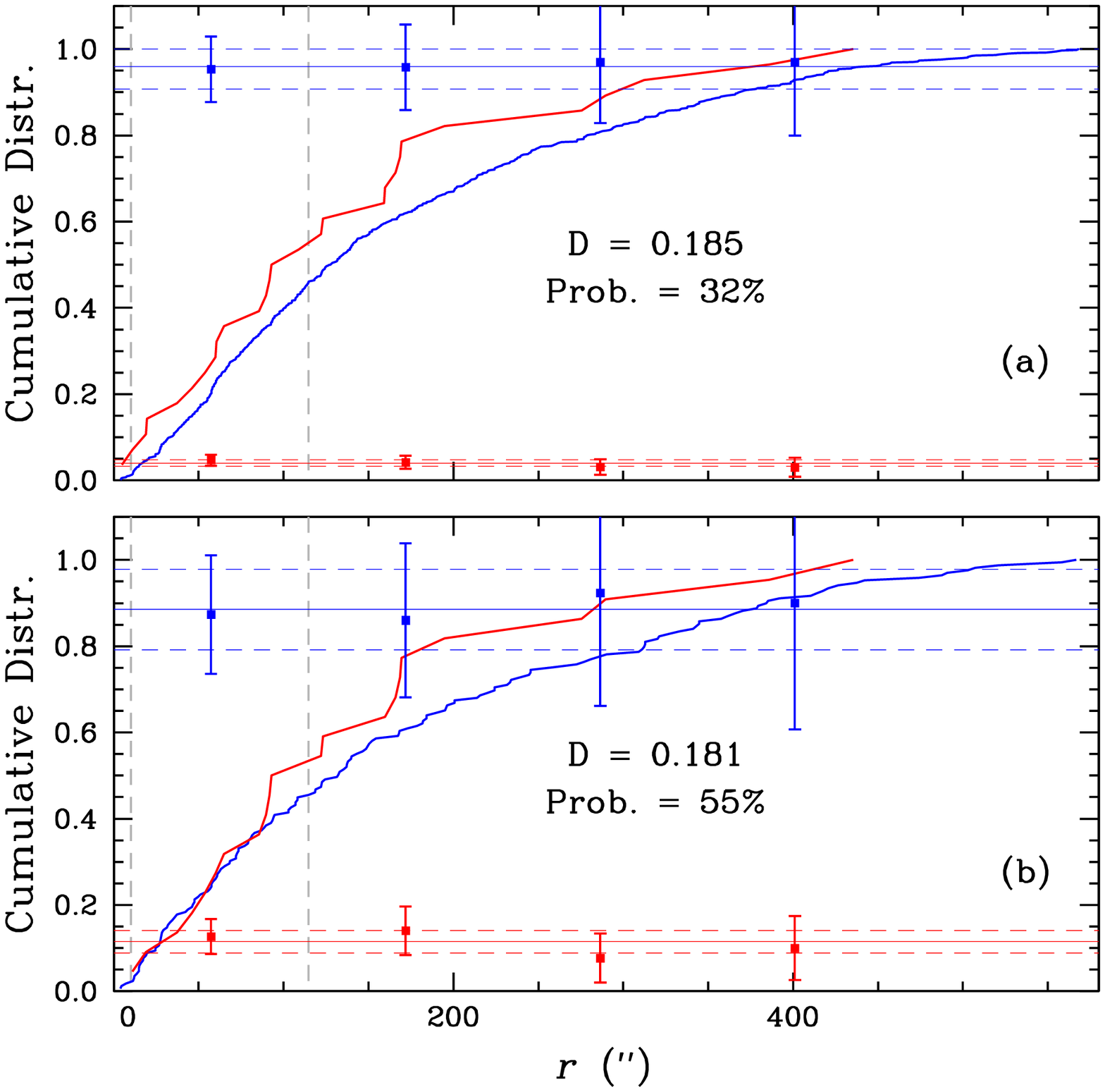}
\caption{(a) Comparison of the cumulative radial distributions of the AGB and the RGB stars.
The red line is for the AGB stars, while the blue line is for the RGB stars. 
The vertical dashed lines are for the core and the half-light radii of the cluster. 
The blue and red horizontal lines are for the number ratios of the AGB and the RGB stars, which do not vary against the radial distance.
(b) Same as (a) but for the \cnw\ AGB and the \cnw\ RGB stars.
}\label{fig:agbrad}
\end{figure}

\subsubsection{Multiple AGB populations in NGC~6752}
It has long been known that AGB stars can have different CN abundance from their progenitors (i.e., RGB stars) in some GCs \citep[e.g., see][]{pilachowski96,sneden00}.
In their pioneering study, \citet{norris81} found that NGC~6752 does not possess the \cns\ AGB population, while the \cns\ RGB stars are a major component of the cluster as already shown in Figure~\ref{fig:cmdnorris}, which led them to propose that the \cns\ RGB stars evolve into the \agbm\ phase.
In Figure~20(a) of \citet{lee17}, we showed a plot of \agbrgb\ versus the HB type of Galactic GCs, which supports the idea that the significance fraction of the HB stars in GCs containing a BHB population only must have evolved into the \agbm\ sequence and failed to reach the AGB sequence.

More recently, \citet{campbell13} claimed that all AGB stars in their sample have low Na abundances, which is strong evidence of the lack of SG AGB stars in NGC~6752 and confirms the previous result by \citet{norris81}.
Later, however, \citet{lapenna16} re-analyzed the same AGB stars that \citet{campbell13} studied and claimed that NGC~6752 contains a significant fraction of an intermediate AGB population with enhanced sodium abundances ([Na/Fe] $\lesssim$ +0.4).\footnote{Note that at least four intermediate AGB stars classified by \citet{lapenna16}, based on the [Na/Fe] abundances, are \cnw\ AGB stars classified by \citet{norris81}, based on the \scn\ measurements; 1620(Lapenna et al.) = CS112(Norris et al), 89 = A63, 94 = A10 and 53=CS1.}
It should be mentioned that \citet{villanova09} found at least a single HB star with sodium enhancement ([Na/Fe] $\approx$ +0.5) in a region cooler than the Grundahl jump,  that will eventually evolve into the AGB sequence.

In Figure~\ref{fig:agbcmd}, we show the \vby\ and the \vcn\ CMDs for all AGB stars that we found in NGC~6752.
Although the double AGB sequences in NGC~6752 are not as distinctive as those in M5 \citep[see Figure~21 of][]{lee17}, NGC~6752 appears to have some extreme \cns\ AGB population with \cnjwlcor\ $>$ 0.0.
However, very unfortunately, these three extreme \cns\ AGB stars classified from our photometry have no previous spectroscopic study.
Future high-resolution spectroscopic studies of these stars are highly desirable to rectify the MSPs of the AGB populations in NGC~6752.
Using the \cnjwlcor\ distribution shown in Figure~\ref{fig:agbcmd}(c), we performed a Hartigan's dip-test to see if the \cnjwlcor\ distribution of the AGB stars is unimodal.
We obtained $D$ = 0.053 and a $p$-value = 0.814, suggesting that it is not unimodal.

We made a comparison of our photometry with elemental abundance measurements by others in order to examine the basis of our AGB classification. 
In Figure~\ref{fig:agbspec}, we show the elemental abundances of the AGB stars by \citet{lapenna16} against \cnjwlcor. In the figure, we also show the RGB stars by \citet{carretta07} and \citet{yong08}.
The figure suggests that at least three AGB stars denoted as red open circles belong to the \cns\ population from our photometry.
On the other hand, nine AGB stars classified as an intermediate population from the sodium abundance by \citet{lapenna16} do not appear to be the \cns\ population from our photometry.
The discrepancy in the elemental abundances is beyond the scope of the current work and we decline to discuss this matter further \citep[see, for example,][]{lee10,lee16,campbell17}.
Our main purpose here is to show that our \cnjwlcor\ is still useful to study multiple AGB populations in GCs.

The number ratio of the AGB stars from our photometry alone becomes \nrgb\ = 79:21 ($\pm$13), which is significantly different from that of RGB stars, \nrgb\ = 25:75 ($\pm$3), suggesting that a significant fraction of the \cns\ population does not evolve through the AGB phase in NGC~6752.
It is very unlikely, but if we include the nine intermediate population AGB stars from \citet{lapenna16} as the \cns\ population, the AGB number ratio becomes marginally consistent with that of RGB stars within the statistical errors, \nrgb\ = 43:57 ($\pm$29).

\begin{figure}
\epsscale{1.0}
\figurenum{20}
\plotone{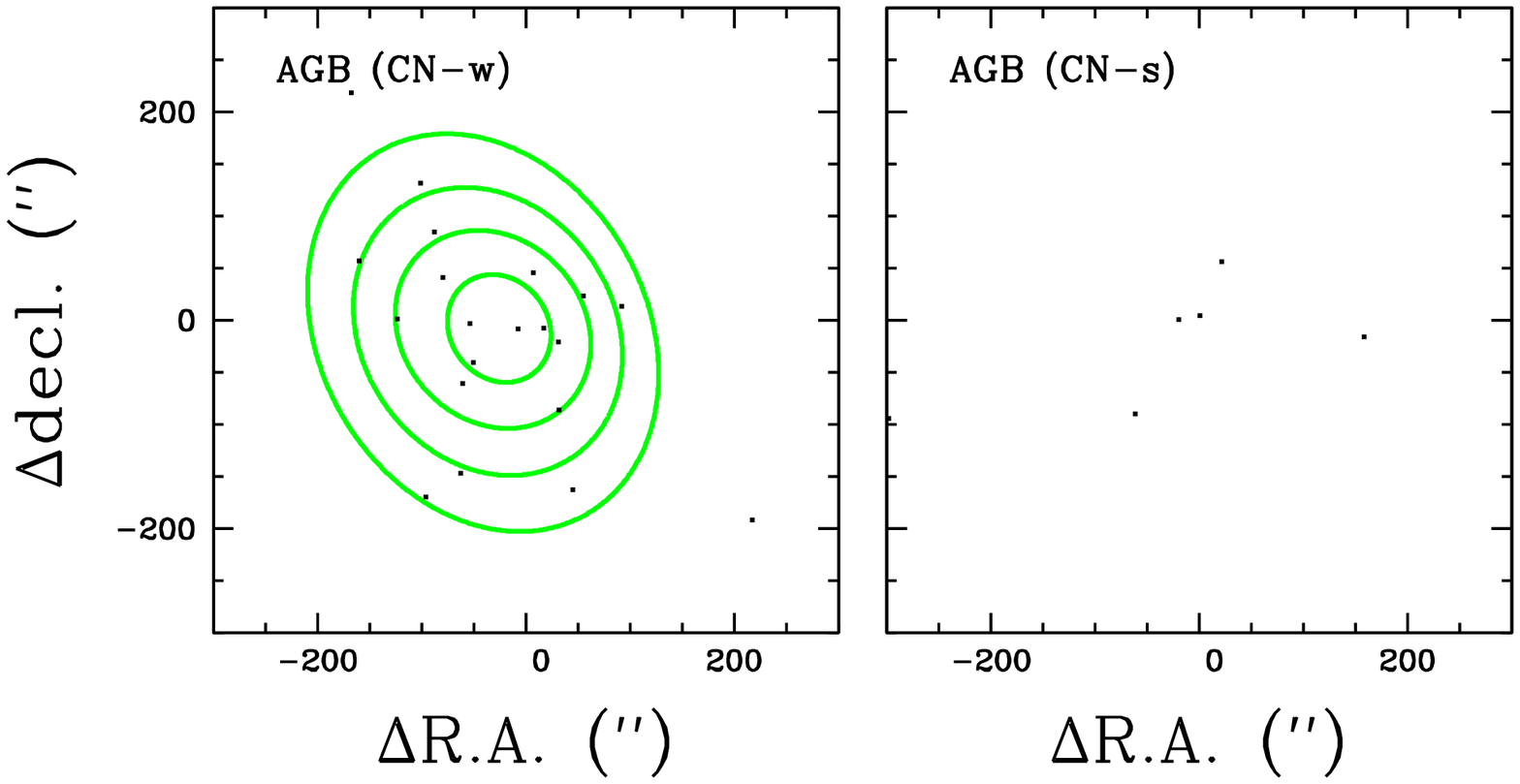}
\caption{Spatial distribution of the AGB stars in NGC~6752.
For the \cnw\ AGB population we show the isodensity contour with green solid lines.
Note that the spatial distribution of the \cnw\ AGB population is more elongated along the NNW--SSE direction, consistent with that of the \cnw\ RGB population.
}\label{fig:agbdistr}
\end{figure}

\subsubsection{Radial and spatial distributions of AGB stars}
In Figure~\ref{fig:agbrad}, we show the comparisons of the cumulative radial distributions of the AGB populations and those of the RGB populations.
As shown, the radial distribution of the AGB stars is very similar to that of the RGB stars.
We performed the K--S test and we found the probability that the both distributions are drawn from an identical population is 32\%, with a K--S discrepancy of 0.185 for all the whole AGB and RGB stars, and the probability is 55\%, with a a K--S discrepancy of 0.181, for the \cnw\ AGB and \cnw\ RGB stars, strongly suggesting that the AGB and RGB stars are most likely drawn from identical parent distributions.

In Figure~\ref{fig:agbdistr}, we show the spatial distribution of the \cnw\ AGB stars.
In the figure, it is evident that the \cnw\ AGB stars are preferentially located along the NNW--SSE direction, consistent with the spatial distribution of the \cnw\ RGB stars, suggesting that both the AGB and the RGB stars share the same structural, and furthermore, dynamical properties in NGC~6752, which would not be a surprise if the \cnw\ RGB population were the progenitor of the \cnw\ AGB population.

\begin{figure}
\epsscale{1.0}
\figurenum{21}
\plotone{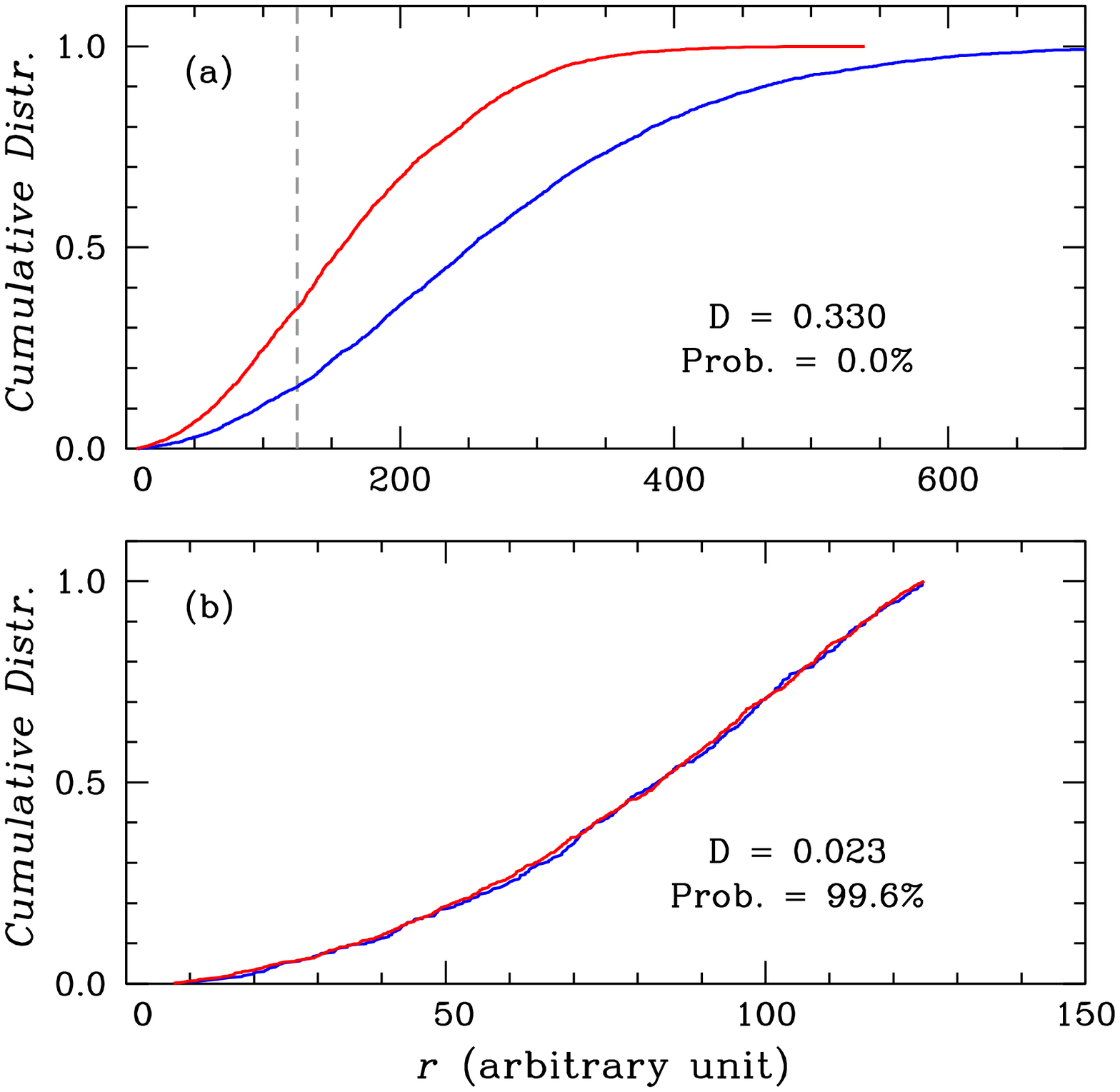}
\caption{(a) Comparison of the Monte-Carlo simulations of the cumulative radial distributions of 2-dimensional Gaussian distributions.
The blue line is for an asymmetric distribution with $\sigma_{\rm X}$ = 300 and $\sigma_{\rm Y}$ = 360 and the red line is for a symmetric distribution with $\sigma_{\rm X}$ = 200 and $\sigma_{\rm Y}$ = 200 (the unit for the radial distance is arbitrary). The vertical dashed gray line denotes the boundary of the truncation at $r_{\rm trunc}$ = 125.
(b) Same as (a), but for the truncated distributions.
A K--S test suggests that these two truncated initially heterogeneous cumulative radial distributions are identical.
}\label{fig:simrad}
\end{figure}

\section{SUMMARY}
We showed that our photometric indices, \cnjwl\ or \cnjwlcor, are as powerful as the traditional spectroscopic indices, \scn\ or \ds, and they can correctly and accurately distinguish the MSPs of cool stars (i.e., RGB and AGB) in a dense stellar environment.
Thanks to the capabilities of our new system, we were able to provide important observational lines of evidence to shed more light on the true nature of MSPs in NGC~6752.

Our new photometry using the \cnjwlcor\ index clearly exhibits the discrete double RGB sequence in NGC~6752, with the number ratio of \nrgb\ = 25:75 ($\pm$ 3), consistent with the previous estimate from the \hst\ photometry by \citet{milone17}.
Furthermore, our photometric nitrogen abundance showed that NGC~6752 most likely has a bimodal nitrogen distribution.

Similar to M5 \citep[e.g., see][]{smith13,lee17}, NGC~6752 exhibits discontinuities in the \cnjwlcor\ versus [O/Fe] and the \cnjwlcor\ versus [Na/Fe] relations (i.e., the discontinuous [N/Fe] versus [O/Fe] and [N/Fe] versus [Na/Fe] relations) between the \cnw\ and the \cns\ RGB populations.
The discontinuous or discrete evolution of the nitrogen abundance, along with the apparently continuous evolution of oxygen and sodium abundances, as can be witnessed in the plot of Na--O anticorrelations in normal GCs, may pose a difficult but critical constraint on the astrophysical sources of these elements.

We explored the RGB bump $V$ magnitudes, finding that the \vbump\ magnitude of the \cns\ population is slightly brighter than that of the \cnw\ population, $\Delta$\vbump = 0.04 $\pm$ 0.04 mag.
With no perceptible spread in age and metallicity between the two populations, the difference in the \vbump\ can be translated into the difference in the helium abundance by \dy\ = 0.016 $\pm$ 0.016, in excellent agreement with previous estimates by  \citet{milone13} and \citet{largioia18}, in the sense that the \cns\ population can be slightly more helium-enhanced. 

Using various methods, we showed that the both RGB populations have almost identical center positions.
Similarly, the cumulative radial distributions of both populations are likely identical, which can also be confirmed by the flat moving average of our \cnjwlcor index,$\langle$\cnjwlcor$\rangle_{\rm 25\ pts}$, and by the consistent SBPs of both populations.
As we have already pointed out for M5, however, the spatial distributions of individual populations may render a very different picture.
The projected spatial distribution of the \cnw\ RGB population is elongated along the NNW--SSE direction.
Moreover, the elongated spatial distribution of the \cnw\ AGB stars is consistent with that of the \cnw\ RGB stars, strongly suggesting that they share the same structural, and perhaps kinematic, properties.
From our calculation of the mean rotations for individual populations, it is believed that the elongated spatial distribution of the \cnw\ population is due to its fast rotation.
It is thought that NGC~6752 is another exemplary GC showing the structure-kinematics coupling.

The differences in the spatial distribution and the projected rotation between the two populations are difficult to understand in the context of the self-enrichment formation scenario of normal GCs.
For example, the numerical simulation by \citet{bekki10} showed that the SG population (i.e., the \cns\ population in the context of the self-enrichment scenario) is supposed to have more elongated substructure with fast rotation.
Compared to its large age, 12.50 $\pm$ 0.25 Gyr \citep{vandenberg13}, the currently believed relaxation time scale at the half-light radius of NGC~6752 is very short, $\approx$ 0.7 Gyr \citep{harris96}.
According to the recent $N$-body numerical simulations \citep[see, for example][]{vesperini13}, it has been claimed that the time scale for achieving the complete mixing may take at least 20 half-mass relaxation time, which is compatible with the age of the cluster.
Therefore, any structural signatures engraved at the epoch of the GC formation are expected to be gradually erased during the course of the long-term dynamical evolution, resulted in the flat number ratios between the two populations.
From this view, the discrepancy between the radial and the spatial distributions of the MSPs in NGC~6752 may pose a somewhat contradictory problem, which was also pointed out for M5 in our previous study \citep{lee17}.
At this point, it is not clear if the homogenization in the cumulative radial distributions between the MSPs works in different time scale than that in the spatial distributions.
The substantial net rotation of the \cnw\ population, while the weak net or no rotation of the \cns\ population could be a strong observational line of constraint to understand the long-term dynamical evolution of the GC system.

Could our discovery here reflect the fact that the cumulative radial distributions of the individual populations are not a proper indicator of the degree of equipartition of kinetic energy?
As demonstrated in Figure~\ref{fig:simrad}, the truncated initially heterogeneous 2-dimensional gaussian distributions can end up with very similar cumulative radial distributions without invoking the dynamical evolution of GCs.
If it is true, the tidal truncation by the Milky Way is most likely responsible for the very similar cumulative  radial distributions between MSPs in NGC~6752 and M5.
On the other hand, the homogenization of the spatial distribution and rotation may have taken a much longer time-scale than is currently believed.
Future work in this direction is needed.

For the first time, we have presented a complete census of the MSPs of the AGB stars in NGC~6752.
We found that NGC~6752 is most likely contains at least a few extreme \cns\ AGB stars from our new photometry, which is in accordance with the existence of a sodium-enhanced HB star redder than the Grundahl jump \citep{villanova09}.
Unfortunately, there is no previous spectroscopic study for these AGB stars and future high-resolution spectroscopic study of these stars is needed to shed more light on the evolution of low-mass stars.

\acknowledgements
J.-W.L. acknowledges financial support from the Basic Science Research Program (grant No. 2016-R1A2B4014741) through the National Research Foundation of Korea (NRF) funded by the Korea government (MSIP).
The author also thanks the anonymous referee for constructive comments.

\end{document}